\documentclass[twocolumn]{aastex62}
\usepackage{subfigure}
\usepackage{ulem}

\graphicspath{{./}{figures/}}

\received{2020 March 11}
\revised{2020 April 19}
\accepted{2020 April 27}
\published{2020 July 10}

\shorttitle{Early Observations of SN\,2019ein}
\shortauthors{Pellegrino et al.}

\begin{document}

\title{Constraining the Source of the High-velocity Ejecta in Type Ia SN\,2019ein}


\author[0000-0002-7472-1279]{C. Pellegrino}
\affil{Las Cumbres Observatory, 6740 Cortona Drive, Suite 102, Goleta, CA 93117-5575, USA}
\affil{Department of Physics, University of California, Santa Barbara, CA 93106-9530, USA}
\author[0000-0003-4253-656X]{D. A. Howell}
\affil{Las Cumbres Observatory, 6740 Cortona Drive, Suite 102, Goleta, CA 93117-5575, USA}
\affil{Department of Physics, University of California, Santa Barbara, CA 93106-9530, USA}
\author[0000-0002-4781-7291]{S. K. Sarbadhicary}
\affil{Department of Physics and Astronomy, Michigan State University, East Lansing, MI 48824, USA}
\author{J. Burke}
\affil{Las Cumbres Observatory, 6740 Cortona Drive, Suite 102, Goleta, CA 93117-5575, USA}
\affil{Department of Physics, University of California, Santa Barbara, CA 93106-9530, USA}
\author[0000-0002-1125-9187]{D. Hiramatsu}
\affil{Las Cumbres Observatory, 6740 Cortona Drive, Suite 102, Goleta, CA 93117-5575, USA}
\affil{Department of Physics, University of California, Santa Barbara, CA 93106-9530, USA}
\author[0000-0001-5807-7893]{C. McCully}
\affil{Las Cumbres Observatory, 6740 Cortona Drive, Suite 102, Goleta, CA 93117-5575, USA}
\affil{Department of Physics, University of California, Santa Barbara, CA 93106-9530, USA}
\author{P. A. Milne}
\affiliation{Department of Astronomy/Steward Observatory, 933 North Cherry Avenue, Rm. N204, Tucson, AZ 85721-0065, USA}
\author[0000-0003-0123-0062]{J. E. Andrews}
\affil{Department of Astronomy/Steward Observatory, 933 North Cherry Avenue, Rm. N204, Tucson, AZ 85721-0065, USA}
\author[0000-0001-6272-5507]{P. Brown}
\affil{Department of Physics and Astronomy, Texas A\&M University, 4242 TAMU, College Station, TX 77843, USA}
\affil{George P. and Cynthia Woods Mitchell Institute for Fundamental Physics \& Astronomy, USA}
\author[0000-0002-8400-3705]{L. Chomiuk}
\affil{Department of Physics and Astronomy, Michigan State University, East Lansing, MI 48824, USA}
\author[0000-0003-1039-2928]{E. Y. Hsiao}
\affil{Department of Physics, Florida State University, Tallahassee, FL 32306, USA}
\author[0000-0003-4102-380X]{D. J. Sand}
\affil{Department of Astronomy/Steward Observatory, 933 North Cherry Avenue, Rm. N204, Tucson, AZ 85721-0065, USA}
\author[0000-0002-9301-5302]{M. Shahbandeh}
\affil{Department of Physics, Florida State University, Tallahassee, FL 32306, USA}
\author[0000-0001-5510-2424]{N. Smith}
\affiliation{Department of Astronomy/Steward Observatory, 933 North Cherry Avenue, Rm. N204, Tucson, AZ 85721-0065, USA}
\author[0000-0001-8818-0795]{S. Valenti}
\affil{Department of Physics, University of California, 1 Shields Avenue, Davis, CA 95616-5270, USA}
\author[0000-0001-8764-7832]{J. Vink{\'o}}
\affil{CSFK Konkoly Observatory, Konkoly-Thege ut 15-17, Budapest, 1121, Hungary}
\affil{Department of Optics and Quantum Electronics, University of Szeged, Dom ter 9, Szeged, 6720, Hungary}
\affil{Department of Astronomy, University of Texas at Austin, 2515 Speedway, Austin, TX, USA}
\author[0000-0003-1349-6538]{J. C. Wheeler}
\affil{Department of Astronomy, University of Texas at Austin, 2515 Speedway, Austin, TX, USA}
\author{S. Wyatt}
\affil{Department of Astronomy/Steward Observatory, 933 North Cherry Avenue, Rm. N204, Tucson, AZ 85721-0065, USA}
\author[0000-0002-6535-8500]{Y. Yang}
\affil{Department of Particle Physics and Astrophysics, Weizmann Institute of Science, Rehovot 76100, Israel}

\begin{abstract}

We present multiwavelength photometric and spectroscopic observations of SN\,2019ein, a high-velocity Type Ia supernova (SN Ia) discovered in the nearby galaxy NGC 5353 with a two-day nondetection limit. SN\,2019ein exhibited some of the highest measured expansion velocities of any SN Ia, with a Si II absorption minimum blueshifted by 24,000 km s$^{-1}$ at 14 days before peak brightness. More unusually, we observed the emission components of the P Cygni profiles to be blueshifted upward of 10,000 km s$^{-1}$ before \textit{B}-band maximum light. This blueshift, among the highest in a sample of 28 other Type Ia supernovae, is greatest at our earliest spectroscopic epoch and subsequently decreases toward maximum light. We discuss possible progenitor systems and explosion mechanisms that could explain these extreme absorption and emission velocities. Radio observations beginning 14 days before \textit{B}-band maximum light yield nondetections at the position of SN\,2019ein, which rules out symbiotic progenitor systems, most models of fast optically thick accretion winds, and optically thin shells of mass $\lesssim 10^{-6}$ M$_\odot$ at radii $< 100$ AU. Comparing our spectra to models and observations of other high-velocity SNe Ia, we find that SN\,2019ein is well fit by a delayed-detonation explosion. We propose that the high emission velocities may be the result of abundance enhancements due to ejecta mixing in an asymmetric explosion, or optical depth effects in the photosphere of the ejecta at early times. These findings may provide evidence for common explosion mechanisms and ejecta geometries among high-velocity SNe Ia.

\end{abstract}

\keywords{supernovae: individual (SN 2019ein)}

\section{Introduction} \label{sec:intro}

Supernovae Ia (SNe Ia) are thermonuclear explosions involving at least one white dwarf (WD) progenitor star \citep{Bloom}. A unique characteristic of SNe Ia is that they show a relationship between their peak luminosity and the width of their light curve, known as the Phillips relation \citep{Phillips}. This correlation allows the calibration of absolute brightness by light curve shape, which enables the determination of distances on cosmological scales. As standardizable candles, observations of SNe Ia have revealed the existence of dark energy (e.g. \citealt{Riess1998,Schmidt1998,Perlmutter1999}) and allow for a low-redshift measurement of the Hubble constant (e.g. \citealt{Riess2019}). A better understanding of their progenitor systems, explosion mechanisms, and observational characteristics is important to mitigate systematic uncertainties in order to use these objects for cosmological measurements.

Over the last several decades, sky surveys and deep imaging have led to the discovery of thousands of SNe Ia (e.g. \citealt{Guy2010,Silverman,Macaulay}). Large samples have shown that significant diversity exists within the population of SNe Ia (e.g. \citealt{Parrent}). Obtaining detailed observations of SNe Ia is important for understanding the sources of this diversity. While the majority of SNe Ia are ``normal'' and obey the Phillips relation, a sizeable minority of peculiar objects tends to show varied photometric and spectral evolution around peak brightness \citep[e.g.][]{Filippenko1992a, Filippenko1992, Phillips1992}, suggesting that fundamental differences beyond luminosity exist in the population of SNe Ia.

Models of progenitor systems and explosion mechanisms have attempted to explain the observed photometric and spectroscopic heterogeneity. Most SN Ia progenitor systems are modeled by accretion onto a degenerate WD from a nondegenerate companion (the single-degenerate scenario, e.g. \citealt{Whelan1973}) or by the accretion or merger of two degenerate WDs (the double-degenerate scenario, e.g. \citealt{Iben1984}). In addition, a variety of theoretical explosion models have been able to reproduce observed characteristics of SNe Ia. One such model is a delayed-detonation explosion, where a (subsonic) deflagration flame transitions to a (supersonic) detonation at some transition density \citep{Iwamoto,Nomoto2013}. Delayed-detonation simulations are able to reproduce a wide variety of light-curve widths, $^{56}$Ni masses, ejecta compositions, and ejecta velocities in Chandrasekhar-mass progenitor WDs \citep{Khokhlov,Seitenzahl}. Another popular model is the double-detonation explosion, in which a detonation of helium accreted onto the surface of a WD leads to a second detonation at the core of the star \citep{Fink2010, Kromer2010, Woosley2011}.

Several observational classification schemes have been proposed that may indirectly probe these different physical models. \citet{Branch2006} propose one such scheme, in which SNe Ia are classified by the strength of their Si II absorption features at maximum light. Additionally, \citet{Wang2009} sort SNe Ia by their Si II velocities, measured from the minimum of the absorption trough at \textit{B}-band maximum light, into two classes: a high-velocity (HV) class, with $v_{\text{Si II}} \gtrsim$ 12,000 km s$^{-1}$, and a normal class, with $v_{\text{Si II}} \lesssim$ 12,000 km s$^{-1}$. After maximum light, SNe Ia can be classified as either high-velocity gradient (HVG) or low-velocity gradient (LVG) if the measured Si II velocity gradient is above or below 70 km s$^{-1}$ day$^{-1}$, respectively \citep{Benetti2005}.

This diversity in velocity may arise from different distributions of Si in the outer layers of the ejecta, which in turn depend on the explosion mechanism. For instance, \citet{Mazzali} studied the Si II and Ca II absorption features in the Type Ia SN\,1999ee and found that two separate components, separated by over 7,000 km s$^{-1}$, were visible in the spectra before \textit{B}-band maximum light. The authors described these as high-velocity features (HVFs) and photospheric velocity features (PVFs) and suggested they could be the result of additional mass at HVs. Other studies have attributed HVFs, particularly of the Ca II NIR feature, to interactions between the SN shock wave and a shell of circumstellar material formed from the SN progenitor system (e.g. \citealt{Gerardy,Mulligan2018,Mulligan2017}). 

One distinguishing feature between explosion mechanisms is the symmetry of the ejecta. \citet{Kasen2007} modeled spectroscopically normal SNe Ia and found that in asymmetric explosions, the color evolution and Si II 6355 \AA{} velocity evolution exhibit significant viewing-angle dependence. Additionally, \citet{Maund} found an empirical relation between the Si II 6355 \AA{} velocity gradient, as originally defined by \citet{Benetti2005}, and the polarization across the same line, which traces the degree of the Si asymmetry in the ejecta (see, e.g., \citealt{WangWheeler} for a review).  Therefore, a better understanding of the spectroscopic differences of SNe Ia is crucial to constraining their explosion mechanisms and ejecta geometries. 

In this paper we present observations of SN\,2019ein, an extreme HV SN Ia. Our early-time observations, beginning two weeks before maximum light, make SN\,2019ein one of the best-studied HV SN Ia. The earliest spectral data at 14 days before \textit{B}-band maximum light reveal some of the highest ejecta velocities ever measured. Perhaps more interestingly, the emission features in the P Cygni profile of SN\,2019ein are blueshifted with respect to the redshift of its host galaxy. This systematic offset is greatest at very early times (several days after explosion) and gradually decreases as the SN evolves. Such a large emission shift sets SN\,2019ein apart from other SNe Ia and hints at a puzzling explosion mechanism and ejecta geometry.  

In Section \ref{sec:methods}, we detail our data acquisition, reduction, and analysis procedure. In Section \ref{sec:phot}, we present comprehensive early-time light curves from the near-ultraviolet (NUV) to the near-infrared (NIR), along with model fits and fitted parameters. In Section \ref{sec:spectra}, we present spectra, measure velocities of absorption features, and compare observations with a delayed-detonation explosion model. In Section \ref{sec:radio}, we place limits on the nature of the progenitor system and the source of HV ejecta using early-time radio observations. In Section \ref{sec:discussion} we offer several possible explanations for the HV ejecta and blueshifted emission features exhibited by SN\,2019ein. Finally, we conclude in Section \ref{sec:conclusions}.

\section{Observations} \label{sec:methods}

\begin{figure}
    \centering
    \includegraphics[scale=0.125]{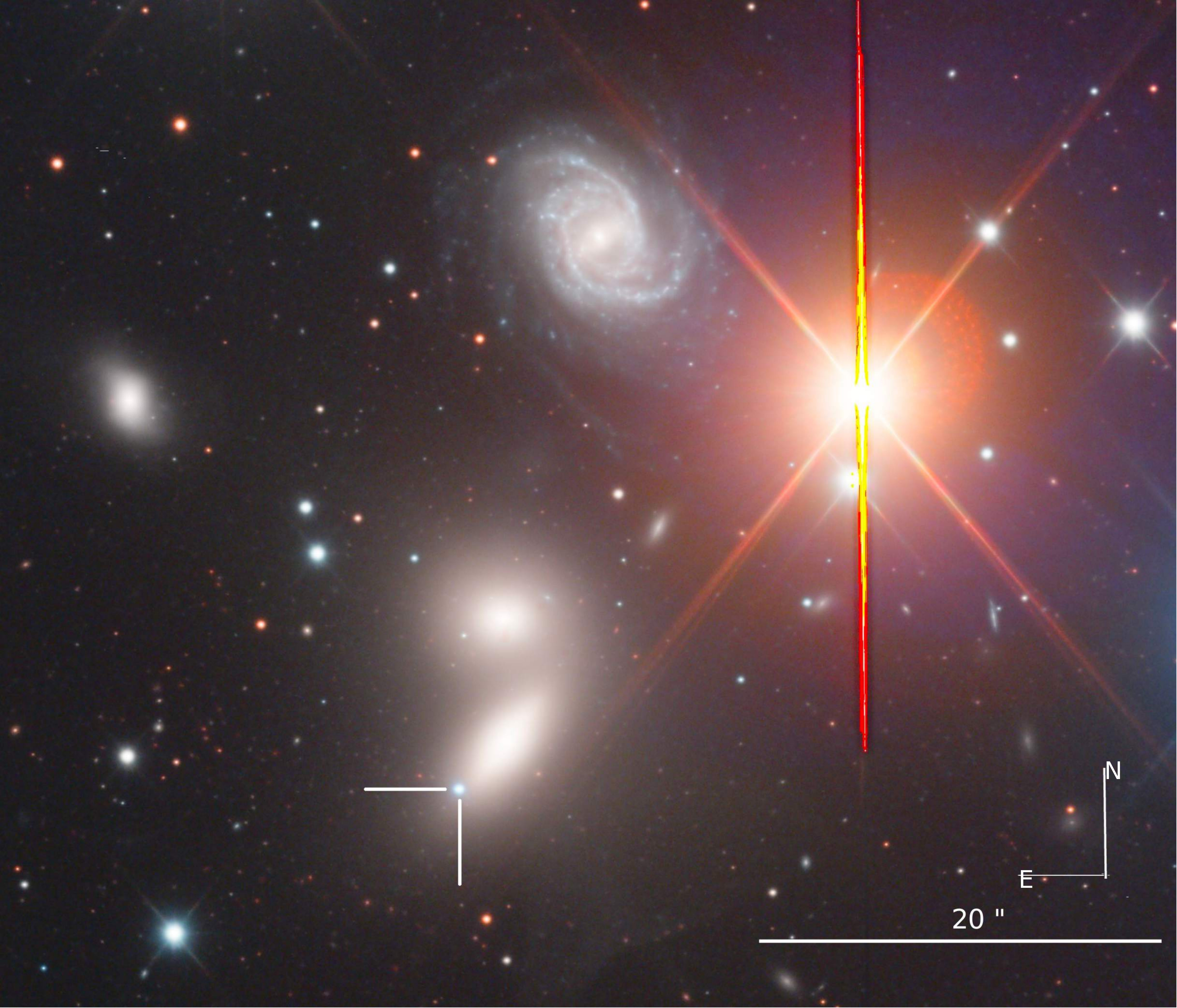}
    \caption{An RGB-color image of SN\,2019ein (shown in the white crosshairs) in NGC 5353 along with its surrounding environment. A scale bar is shown in the bottom right corner. The image was produced using LCO 1m data files courtesy of Peter Il\'a\v{s}.}
    \label{fig:field}
\end{figure}

SN\,2019ein was discovered on MJD 58604.47 (2019 May 1.47) by the ATLAS survey at magnitude 18.194 in their cyan filter \citep{Tonry}. The last nondetection of the transient was by the Zwicky Transient Facility \citep{Bellm} in \textit{r} band at a limit of 19.72 mag on MJD 58602.27 (2019 April 29.27), implying that SN\,2019ein was discovered within two days after explosion. SN\,2019ein exploded in the outskirts of NGC 5353, a lenticular galaxy in a nearby galaxy group. An image of NGC 5353 along with SN\,2019ein is shown in Figure \ref{fig:field}. Using surface brightness fluctuation measurements, J. Jensen et al. (2020, in preparation) measure the distance to NGC 5353 to be 32.96 $\pm$ 1.68 Mpc, and the redshift taken from the NASA/IPAC Extragalactic Database\footnote{https://ned.ipac.caltech.edu} is 0.00775. 

We began daily photometric and spectroscopic follow-up observations of SN\,2019ein starting on 2019 May 2 with the Global Supernova Project using Las Cumbres Observatory (LCO; \citealt{Brown}). Our first spectrum, obtained with the FLOYDS spectrograph on the 2m telescope at Haleakal$\bar{\text{a}}$, allowed LCO to classify SN\,2019ein as a young SN Ia \citep{Burke}. \textit{UBgVri}-band data were obtained with the SBIG, Sinistro, and Spectral cameras on 0.4m, 1m, and 2m telescopes, respectively. With the \texttt{PyRAF}-based photometric reduction pipeline \texttt{lcogtsnpipe} \citep{Valenti2016}, PSF fitting was performed \citep{Stetson}. \textit{UBV}-band photometry was calibrated to Vega magnitudes using Landolt standard fields \citep{Landolt}, while \textit{gri}-band photometry were calibrated to AB magnitudes using the Sloan Digital Sky Survey (SDSS; \citealt{SDSS}). Additionally, magnitudes were corrected for color terms using these standards. Because the SN was offset from the host galaxy, image subtraction was not necessary.  

We obtained four epochs of NIR photometry in \textit{JHK$_{s}$} filters with the 2MASS camera on the Minnesota-60$\arcsec$ telescope on Mt. Lemmon, AZ, as part of the Arizona Transient Exploration and Characterization (AZTEC) program. The data were reduced and stacked with the IRAF\footnote{IRAF is distributed by the National Optical Astronomy Observatories, which are operated by the Association of Universities for Research in Astronomy, Inc., under cooperative agreement with the National Science Foundation.} \texttt{-xdimsum} package. Aperture photometry was obtained with IRAF and calibrated to 20 local standards from the 2MASS catalog \citep{Skrutskie}.

We requested observations from the Ultraviolet and Optical Telescope (UVOT) on the Neil Gehrels Swift Observatory \citep{Gehrels} after the detection of SN\,2019ein. The first epoch of Swift data was obtained on MJD 58605.404 (2019 May 2.4), coincident with the first LCO photometric and spectroscopic epochs. Data were obtained in \textit{uvw2, uvm2, uvw1, u, b,} and \textit{v} filters and reduced using the data-reduction pipeline for the Swift Optical/Ultraviolet Supernova Archive (SOUSA; \citealt{Brown2014}), including applying aperture corrections and zero-points from \cite{Breeveld}. Galaxy subtraction was not performed.

\begin{deluxetable*}{lrrrrr}
\tablenum{1}
\tablecaption{Log of Spectroscopic Observations of SN\,2019ein\label{tab:spec}}
\tablehead{
\colhead{MJD} & \colhead{Phase$^{a}$} & \colhead{Wavelength Range [\AA{}]} & \colhead{Telescope} & \colhead{Instrument}}
\startdata
58605.3 & -14 & 3200 -- 10000 & LCO 2m & FLOYDS \\
58609.5 & -10 & 3500 -- 10000 & LCO 2m & FLOYDS \\
58613.4 & -6 & 3500 -- 10000 & LCO 2m & FLOYDS \\
58615.3 & -4 & 3500 -- 10000 & LCO 2m & FLOYDS \\
58618.2 & -1 & 3500 -- 10000 & LCO 2m & FLOYDS \\
58619.2 & 0 & 3500 -- 10000 & LCO 2m & FLOYDS \\
58628.2 & +9 & 3700 -- 8000 & Bok & BCSpec \\
58628.4 & +9 & 3500 -- 10000 & LCO 2m & FLOYDS \\
58629.2 & +10 & 3700 -- 8000 & Bok & BCSpec \\
58632.4 & +13 & 3500 -- 10000 & LCO 2m & FLOYDS \\
58635.4 & +16 & 3500 -- 10000 & LCO 2m & FLOYDS \\
58638.3 & +19 & 5693 -- 7000 & MMT & Blue Channel \\
58640.3 & +21 & 3700 -- 8000 & Bok & BCSpec \\
58641.4 & +22 & 3500 -- 10000 & LCO 2m & FLOYDS \\
58647.3 & +28 & 3500 -- 10000 & LCO 2m & FLOYDS \\
58651.3 & +32 & 6875 -- 25412 & IRTF & SPeX \\
58653.4 & +34 & 3500 -- 10000 & LCO 2m & FLOYDS \\
58666.4 & +47 & 3500 -- 10000 & LCO 2m & FLOYDS \\
58679.3 & +60 & 3500 -- 10000 & LCO 2m & FLOYDS
\enddata
\tablecomments{\textit{a}: Days relative to \textit{B}-band maximum light}
\end{deluxetable*}

Spectroscopic observations are detailed in Table \ref{tab:spec}. 14 LCO spectra were obtained using the FLOYDS instruments on LCO 2m telescopes at Siding Springs and Haleakal$\bar{\text{a}}$ between -14 days to 60 days with respect to \textit{B}-band maximum. Our spectra cover approximately the entire optical range from 3500 to 10000 \AA{} at resolution R $\approx$ 300-600. Data were reduced using the \texttt{floydsspec} custom pipeline, which performs flux and wavelength calibration, cosmic-ray removal, and spectrum extraction\footnote{https://github.com/svalenti/FLOYDS\_pipeline/blob/master/ \ bin/floydsspec}. In addition, we obtained several spectra in the optical and NIR using the B\&C Spectrograph on the Bok 90$\arcsec$ telescope, the Blue Channel Spectrograph on the MMT at the Fred Lawrence Whipple Observatory, and the SpeX spectrograph \citep{Rayner} in PRISM mode with a 0.5 $\times$ 15" slit on the NASA Infrared Telescope Facility, which was obtained and reduced following the methods in \citet{Hsiao2019}. These data are presented in Section \ref{sec:spectra}. 

\subsection{Radio Observations}\label{sec:radioobs}

Radio observations of SN\,2019ein were obtained with the Karl G. Jansky Very Large Array (VLA) on 2019 May 3 within two days of discovery. Two follow-up observations about a week apart were subsequently obtained.  Each observation was 1 hr long, with 37.6 minutes time on source per block for SN\,2019ein. All observations were taken in \textit{C} band (4-8 GHz) in the \textit{B} configuration (program: 19A-010, PI: L. Chomiuk). The observations were obtained in wide-band continuum mode, yielding 4 GHz of bandwidth sampled by 32 spectral windows, each 64 MHz wide sampled by 2 MHz wide channels. We used 3C286 as our flux and bandpass calibrator, and J1419+3821 as our phase calibrator. Table \ref{tab:radio} contains details of the observations.

\begin{deluxetable*}{lrrrrr}
\tablenum{2}
\tablecaption{Summary of VLA Observations.\label{tab:radio}}
\tablehead{
\colhead{Epoch} & \colhead{Days Since} & \colhead{Synthesized Beam} & \colhead{3$\sigma$-upper Limit$^{c}$}
\\
\colhead{(MJD$^{a}$)} & \colhead{Explosion$^{b}$} & \colhead{(arcsec$\times$arcsec)} & \colhead{($\mu$Jy/beam)}}
\startdata
58606.60 & 3.87 & 1.28$\times$1.27 & 17.88 \\
58614.29 & 11.57 & 2.20$\times$1.23 & 25.32 \\
58620.31 & 17.58 & 1.77$\times$1.99 & 23.46
\enddata
\tablecomments{\\ \textit{a}: MJD at end of observation of each scheduling block\\
\textit{b}: Assuming the explosion happened at most two days before discovery (Section \ref{sec:methods}). The explosion date corresponds to -16.93 days in units of phase relative to \textit{B}-band maximum.\\
\textit{c}: Three times the RMS noise at the site of SN\,2019ein, inside a region of 6$^{\prime\prime}$ radius}
\end{deluxetable*}

We obtained the data sets processed by the VLA CASA calibration pipeline, run on CASA version 5.4.1.\footnote{https://science.nrao.edu/facilities/vla/data-processing/ \ pipeline} The pipeline consists of a collection of algorithms that automatically loads the raw data into a CASA measurement set (MS) format, flags corrupted data (e.g. due to antenna shadowing, channel edges, and radio frequency interference or RFI), applies various corrections (e.g. antenna position and atmospheric opacity) and derives delay, flux-scale, bandpass, and phase calibrations that are applied to the data. 

For each epoch, the \textit{C}-band data were split into 4-6 GHz and 6-8 GHz data sets, and each one was imaged using the CASA routine \texttt{tclean}. We use Briggs weighting of the data with a \texttt{robust=0.7} to provide reasonable balance of angular resolution and source sensitivity. We used multiterm, multifrequency synthesis as our deconvolution algorithm (set with \texttt{deconvolver=`mtmfs'} in \texttt{tclean}), which performs deconvolution on a Taylor-series expansion of the wide-band spectral data in order to minimize frequency-dependent artifacts \citep{Rau2011}. We set \texttt{nterms=2} which uses the first two Taylor terms to create images of intensity and spectral index. Multiple bright radio sources appear off-center in the 8.4$^{\prime}$ field of view, so we use ``w-projection'' (applied with \texttt{gridder=`wproject'} in \texttt{tclean}) to account for non-coplanar effects when deconvolving these sources \citep{Cornwell2008}. The radio nucleus of the host galaxy is the brightest radio source in the field (peak flux $\sim$ 25 mJy) and forms artifacts near the site of the SN, so we performed a phase-only self-calibration with a solution interval of 2 minutes to further clean and reduce the RMS noise in the image. The cleaned and self-calibrated 6-8 GHz image was then convolved to the resolution of the 4-6 GHz image using \texttt{CONVL} in AIPS. Both images were then combined using \texttt{COMB} in AIPS, weighted by their respective RMS noise, to create the final \textit{C}-band image (central frequency of 6 GHz) of the SN\,2019ein field.

No radio source was detected at the site of SN\,2019ein in any of the cleaned deconvolved images down to 3$\sigma$ limits of $\sim$ 18 $\mu$Jy in the first image, and 25 and 23 $\mu$Jy in the subsequent images. We discuss the constraints on progenitor models set by these limits in Section \ref{sec:radio}.

\begin{figure*}
    \centering
    \includegraphics[scale=0.67]{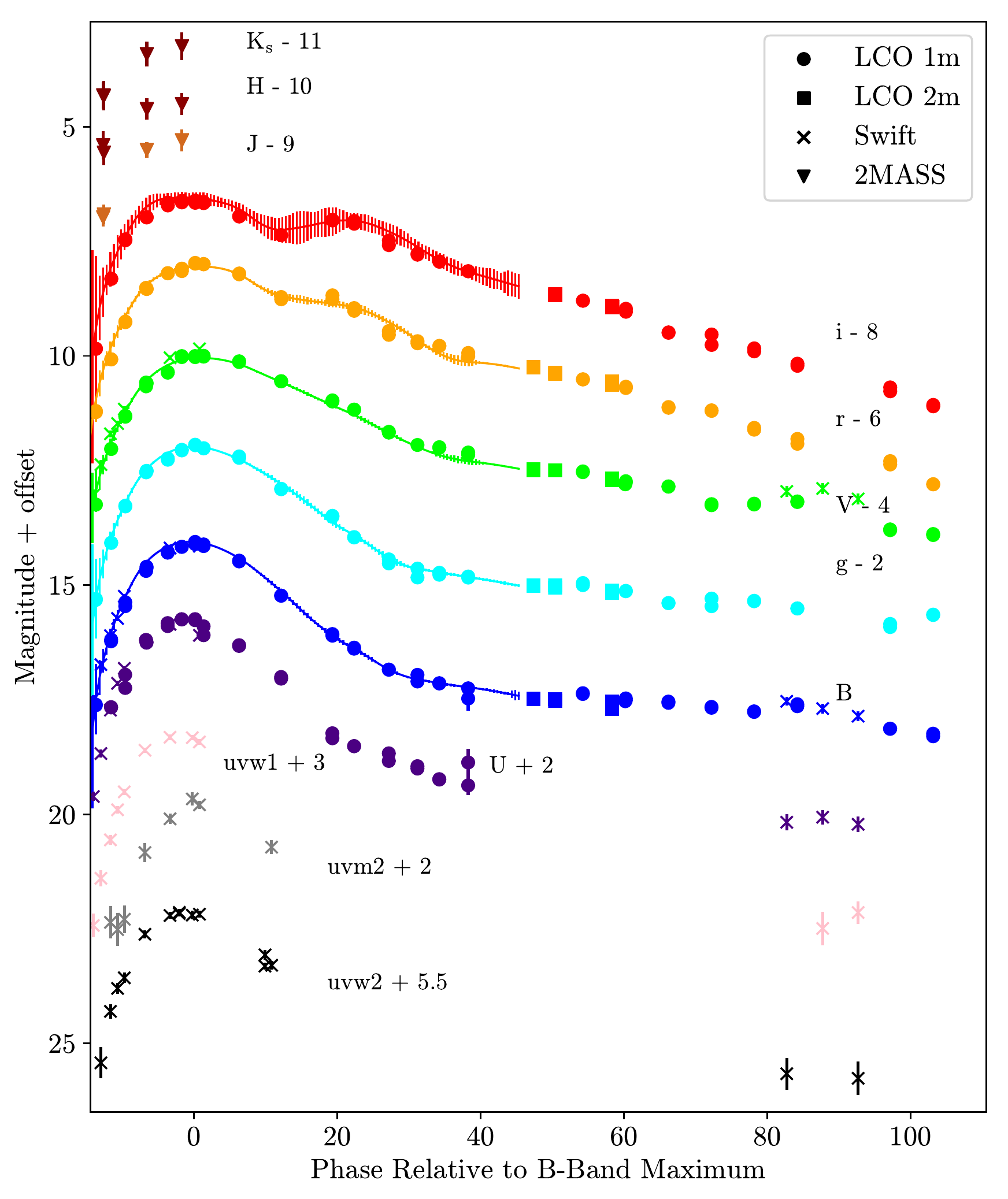}
    \caption{The NUV, optical, and NIR light curves of SN\,2019ein, along with \texttt{SALT2} fits and error bars to LCO data in \textit{BgVri} filters (solid lines). The LCO \textit{UBgVri} photometry is available as Data behind the Figure.}
    \label{fig:lc}
\end{figure*}

\section{Photometric Results} \label{sec:phot}

\subsection{Light Curves of SN\,2019ein} \label{subsec:LCs}

Swift \textit{uvw2, uvm2, uvw1}, Johnson-Cousins \textit{UBV}, SDSS \textit{gri}, and 2MASS \textit{JHK$_{s}$} light curves are shown in Figure \ref{fig:lc}, along with \texttt{SALT2} \citep{Guy} fits to \textit{BgVri} data between -14 days and 40 days with respect to \textit{B}-band maximum light. Our high-cadence observations make the rise of this light curve extremely well-sampled. Because SN\,2019ein was discovered quite early, we are able to tightly constrain the rise time and explosion time. Given the \texttt{SALT2} fits to the light curve, we find the \textit{B}-band maximum occurred at MJD 58619.45 $\pm$ 0.03 (2019 May 16.5), which implies a rise time of $\approx 14$ days since the beginning of observations and a maximum of $17$ days since explosion (all phases hereafter are given in terms of \textit{B}-band maximum light). Using an expanding fireball model, \citet{Kawabata} estimated the explosion time of SN\,2019ein to be MJD 58602.87 $\pm$ 0.55, giving a \textit{B}-band rise time of $\approx$ 16.5 days, which is consistent with our estimates. This fast rise supports the suggestion that HV SNe tend to have shorter \textit{B}-band rise times \citep{Ganeshalingam}.

\texttt{SALT2} fitted parameters are given in Table \ref{tab:params} along with calculated values of absolute magnitude \textit{M}, $\Delta m_{15}(B)$, Milky Way $E(B-V)$, and distance modulus $\mu$. We correct for host galaxy reddening by adopting the value presented in \citet{Kawabata}, who estimate the host extinction as $E(B-V)_{host} = 0.09 \pm 0.02$ mag. Additionally, we use our \texttt{SALT2} parameters to calculate a distance modulus of $\mu = 32.60 \pm 0.07$ \citep{Betoule}, which matches our measured distance modulus from J. Jensen et al. (2020, in preparation). Overall, the fitted parameters show that SN\,2019ein is a photometrically normal SN Ia, albeit with a slightly lower absolute magnitude at peak brightness ($M_{B_{max}} = -18.81 \pm 0.059)$ than expected. For a decline rate of $\Delta m_{15}(B) = 1.40 \pm 0.004$, SNe Ia have on average $M_{B_{max}} \approx -19$ \citep{Hamuy}. Therefore SN\,2019ein falls slightly below the average, even with the modest reddening correction. We find good agreement between our estimated parameters and those derived in \cite{Kawabata}, although our peak \textit{B}-band absolute magnitude is fainter than their estimates, perhaps due to our use of a different distance modulus. Our photometry data are presented in Tables \ref{tab:optphot}, \ref{tab:swiftphot}, and \ref{tab:nirphot}.

\begin{deluxetable}{ccCrlc}[b!] 
\tablenum{3}
\tablecaption{SN\,2019ein Photometric Parameters\label{tab:params}}
\tablecolumns{3}
\tablewidth{0pt}
\tablehead{
\colhead{Parameter} &
\colhead{Value} &
\colhead{Uncertainty} &
}
\startdata
R.A.             & 13:53:29.13 & - \\
Decl.             & +40:16:31.3 & - \\
$x_0$ $^a$             & 0.044   & \pm 0.0007\\
$x_1$ $^a$             & -1.678  & \pm 0.0260\\
$C$ $^a$               & 0.003    & \pm 0.0174 \\
$t_{B max}$ (MJD)$^a$ & 58619.45 & \pm 0.031\\
$\Delta m_{15}(B)^a$   & 1.40    & \pm 0.004 \\
$M_{B_{max}}$ $^b$     & -18.81  & \pm 0.059 \\
$E(B-V)_{MW}$ $^c$   & 0.011 & - \\
$\mu$ $^{d}$           & 32.59 & \pm 0.11 
\enddata
\tablecomments{\\ \textit{a}: From \texttt{SALT2} fits \citep{Guy}\\
\textit{b}: Calculated from \texttt{SALT2} parameters \citep{Betoule}\\
\textit{c}: From \citet{Schlafly}\\
\textit{d}: From J. Jensen et al. (2020, in preparation)\\} 
\end{deluxetable}

\subsection{Color} \label{subsec:color}

\begin{figure}
    \centering
    \includegraphics[scale=0.6]{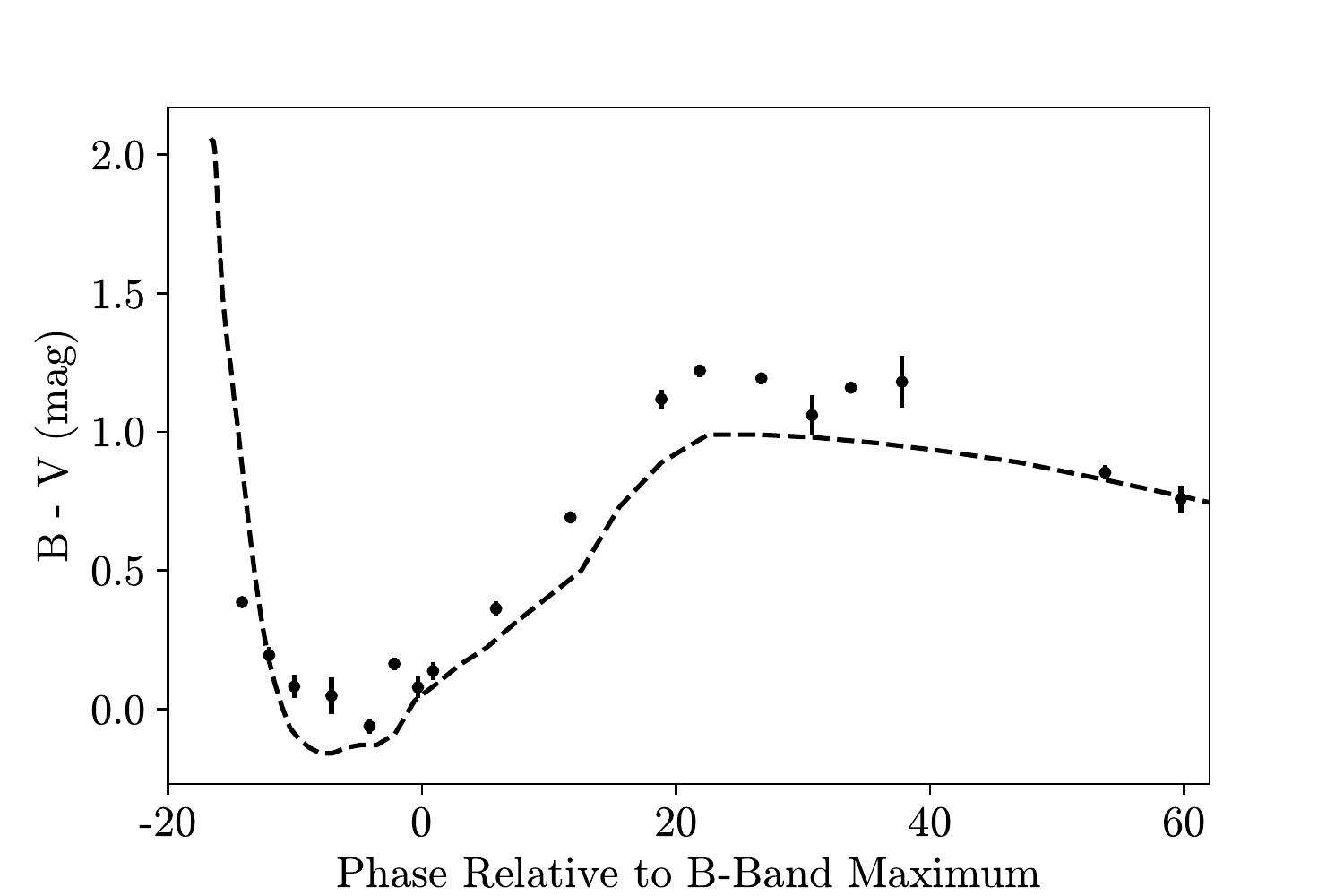}
    \caption{\textit{B-V} color data of SN\,2019ein compared with the color curve of the B15 model of \citet{Blondin2015} for the HV SN\,2002bo (dashed line). Only LCO 1m photometry data obtained before +60 days are presented. For clarity, observations taken on the same day have been median combined. The data have been corrected for Milky Way extinction; neither the data nor the model have been corrected for host galaxy extinction.}
    \label{fig:color}
\end{figure}

The \textit{B-V} color evolution of SN\,2019ein is plotted in Figure \ref{fig:color}, along with the color curve of the delayed-detonation explosion model of \citet{Blondin2015}, hereafter the B15 model. The model broadly matches the data at all phases, particularly around \textit{B}-band maximum, although it tends to predict a bluer color at later phases. Similar trends can be seen in comparisons of other HV SNe with both the B15 model \citep{Gutierrez} and NV SNe \citep{Wang2009}. At early times, the B-V color evolution matches the red group of \citet{Stritzinger2018}, although among this sample SN\,2019ein has a unique Branch classification in this sample \citep{Branch2006}, as described in Section \ref{subsec:branch}. After correcting for host reddening, the \textit{B-V} color of SN\,2019ein is 0.08 $\pm$ 0.04 around \textit{B}-band maximum. This value falls in the overlap between the Normal and HV subsamples of \citet{Foley}.

Additionally, we measure the NUV-optical colors using our Swift photometry. The \textit{uvw1-v} and \textit{u-v} colors one day after \textit{B}-band maximum are 1.58 $\pm$ 0.08 and 0.25 $\pm$ 0.06, respectively. These colors place SN\,2019ein in the NUVR group of \cite{Milne2013}, which is the group of most normal SNe Ia with \textit{u-v} $<$-0.4 at maximum light. This is consistent with results that show HV and HVG SNe are all members of the NUVR group \citep{Milne2013,Brown2018}. Given our velocity measurements discussed in Section \ref{sec:spectra}, the colors of SN\,2019ein fit those of other HV SNe well.

\subsection{Bolometric Luminosity and $^{56}$Ni Mass} \label{subsec:bollum}

Using our maximum-light photometry, we estimate the bolometric maximum luminosity and the corresponding $^{56}$Ni mass. We follow the methods outlined in \citet{Howell2009}: first, the measured flux at \textit{B}-band maximum is calculated by integrating the magnitudes in \textit{U}, \textit{g}, \textit{r}, and \textit{i} filters to ensure optimal wavelength coverage of the optical region. In order to calculate the flux across the rest of the spectrum, we adopted the synthetic spectrum produced by the B15 model. As we show in Section \ref{subsec:explosion}, this model produces close fits to the spectrum of SN\,2019ein at maximum light. We scale the B15 spectrum flux to match the distance of SN\,2019ein. Next we scale the flux of the synthetic spectrum within our filter wavelength ranges to match the observed flux. We then divide this ``warped" flux by the ratio of the measured flux to the total flux in the synthetic spectrum, and define this quantity to be the bolometric flux. 

Following this procedure, we find a maximum bolometric luminosity of $L \sim 7.28 \times 10^{42}$ erg s$^{-1}$. Using the relationship for the luminosity per $^{56}$Ni mass $\dot{S}$ from \citet{Howell2009}, 
\begin{equation}
    \dot{S} = 6.31\times 10^{43}e^{t_r/8.8} + 1.43\times 10^{43}e^{t_r/111} \text{ erg s$^{-1}$ M$\odot^{-1}$}
\end{equation} which is based on Arnett's rule \citep{Arnett}, we calculate a $^{56}$Ni mass of $\sim 0.33$ M$_\odot$ assuming a rise time of $\approx 16.5$ days. This mass is on the low end for SNe Ia \citep{Stritzinger2006},but is supported by the analytic relationship found in \citet{Konyves-Toth} between light-curve width and $^{56}$Ni mass. Because SN\,2019ein is a relatively fast decliner with $\Delta m_{15}(B) = $ 1.40 and is slightly subluminous ($M_B = $ -18.81), we conclude that this $^{56}$Ni mass is a reasonable estimate.

\begin{figure*}
    \includegraphics[width=0.95\textwidth]{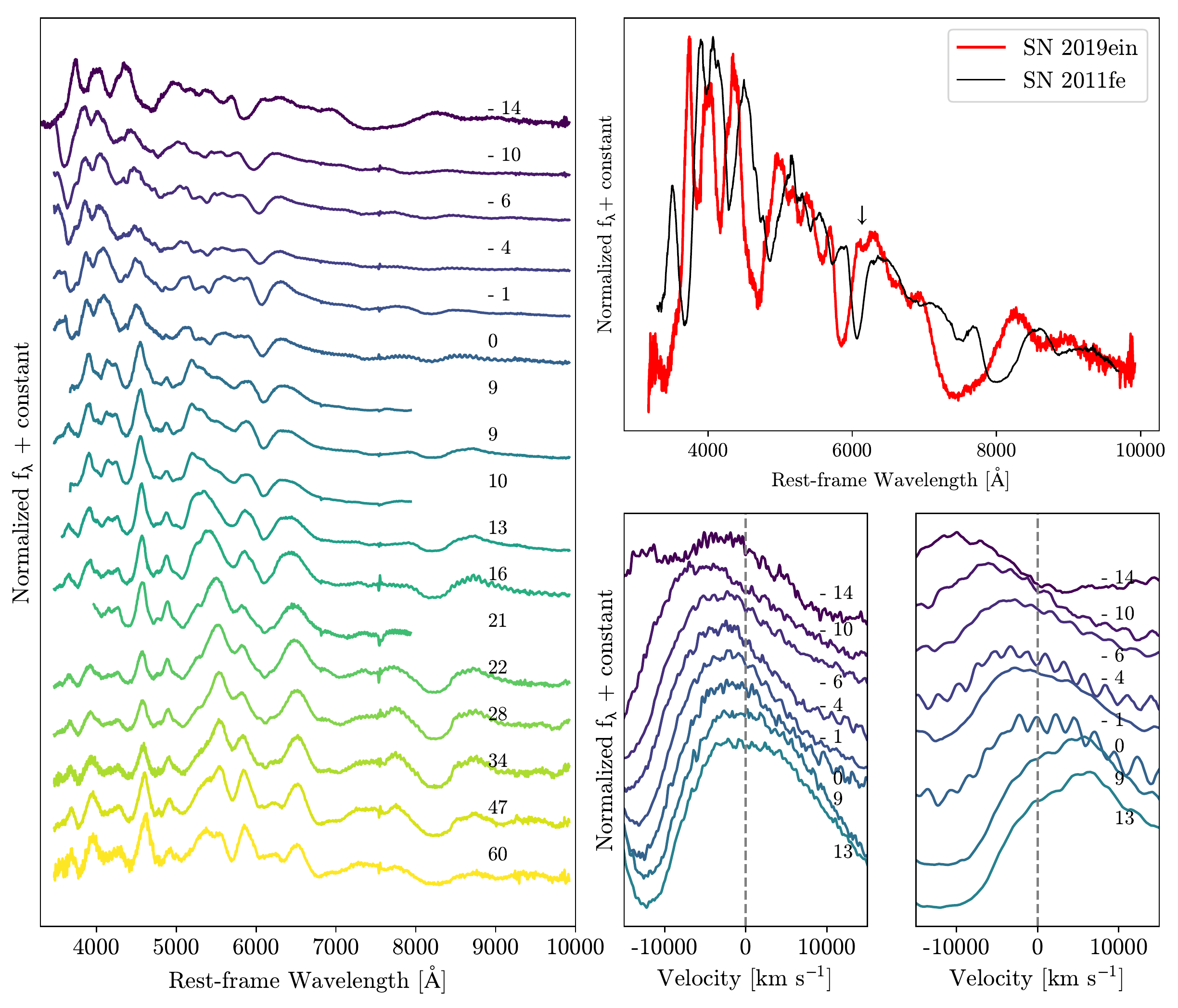}
    \caption{Left: the spectral evolution of SN\,2019ein, from discovery 14 days before \textit{B}-band maximum light until 60 days after \textit{B} maximum. Spectra are described in Table \ref{tab:spec}. The phase with respect to \textit{B}-band maximum is shown to the right of each spectrum. All fluxes are plotted on a linear scale. These spectra are available as Data behind the Figure. Top right: the spectrum of SN\,2019ein at -14 days (red) compared with that of SN\,2011fe at the same phase (black). A downward arrow denotes C II absorption in the spectrum of SN\,2019ein. Bottom right: the emission components of the Si II 6355 \AA{} (left) and Ca II NIR (right) P Cygni profiles from -14 days to 13 days with respect to \textit{B}-band maximum light. The rest wavelengths of these features are shown with dashed lines. We caution that the apparent redshift of the Ca II NIR emission component after maximum light is most likely due to line overlap.}
    \label{fig:allspectra}
\end{figure*}

\section{Spectroscopic Analysis} \label{sec:spectra}

Figure \ref{fig:allspectra} (left) shows the spectral evolution of SN\,2019ein, from -14 to 60 rest-frame days with respect to \textit{B}-band maximum light. In our earliest spectrum (Figure \ref{fig:allspectra}, top right) the most striking features include the broad absorption trough centered at approximately 7500 \AA{}, which is most likely the result of blended Ca II and O I absorption at HVs ($>$ 30,000 km s$^{-1}$), as well as the broad Si II absorption centered at a wavelength less than 6000 \AA{}. The Si II absorption minimum corresponds to a velocity of approximately 24,000 km s$^{-1}$, which is one of the highest velocities ever measured in a SN Ia \citep{Gutierrez}. Additionally, the Ca II H\&K absorption feature is not well defined in this spectrum. This could be due to blending with other absorption lines, or it may be that the line is blueshifted outside of the sensitivity of our spectrograph, although such a blueshift would correspond to a seemingly unphysical velocity of $\sim$45,000 km s$^{-1}$. At this phase, the entire spectrum of SN\,2019ein is noticeably blueshifted with respect to that of SN\,2011fe. Before maximum light the blueshifts of the emission peaks remain prominent; the shifts are greatest in our first epoch, where both the Si II 6355 \AA{} and Ca II NIR components are displaced with velocities upward of 10,000 km s$^{-1}$ (Figure \ref{fig:allspectra}, bottom right). 

Also seen in the earliest spectrum is a small absorption notch, denoted with a black arrow, at a rest-frame wavelength of approximately 6150 \AA{}. This feature is most likely C II 6580 \AA{} at $\approx$ 20,000 km s$^{-1}$, as there is also a possible absorption feature from the C II 7235 \AA{} line at the same velocity. Unburnt carbon in early-time spectra of SNe Ia is not unusual (e.g. \citealt{Parrent2011,Blondin2012,Folatelli2012,Silverman2012,Maguire}); however, few SNe Ia show Si II 6355 \AA{} absorption velocities higher than C II 6580 \AA{} absorption velocities at early times \citep{Parrent2011,Folatelli2012,Silverman2012}, with a notable exception being SN\,2011fe \citep{Parrent2012,Pereira}. The fact that SN\,2019ein shows the opposite trend at this phase may shed light on the explosion mechanism and ejecta geometry. \citet{Parrent2011} note that $v_{\text{C II}}$/$v_{\text{Si II}} < 1$ if the C II feature comes from an asymmetric ejecta distribution viewed at an angle with respect to the observer's line of sight. 

Close to \textit{B}-band maximum, we note a possible HVF in the Ca II H\&K line, as a weaker, lower velocity component becomes visible at roughly -4 days. However, HVFs usually develop at earlier phases, and this feature we observe is equally well fit by Si II absorption, making identification of HVFs at this phase difficult. Except for this exception, we do not find evidence for HVFs in the spectra of SN\,2019ein, and all the velocities we report here are measured from the center of the dominant absorption feature for each line. The lack of two distinct absorption components sets SN\,2019ein apart from most other HV SNe Ia. We discuss possible reasons for this difference in Section \ref{sec:discussion}. 

\begin{figure}
    \centering
    \includegraphics[scale=0.6]{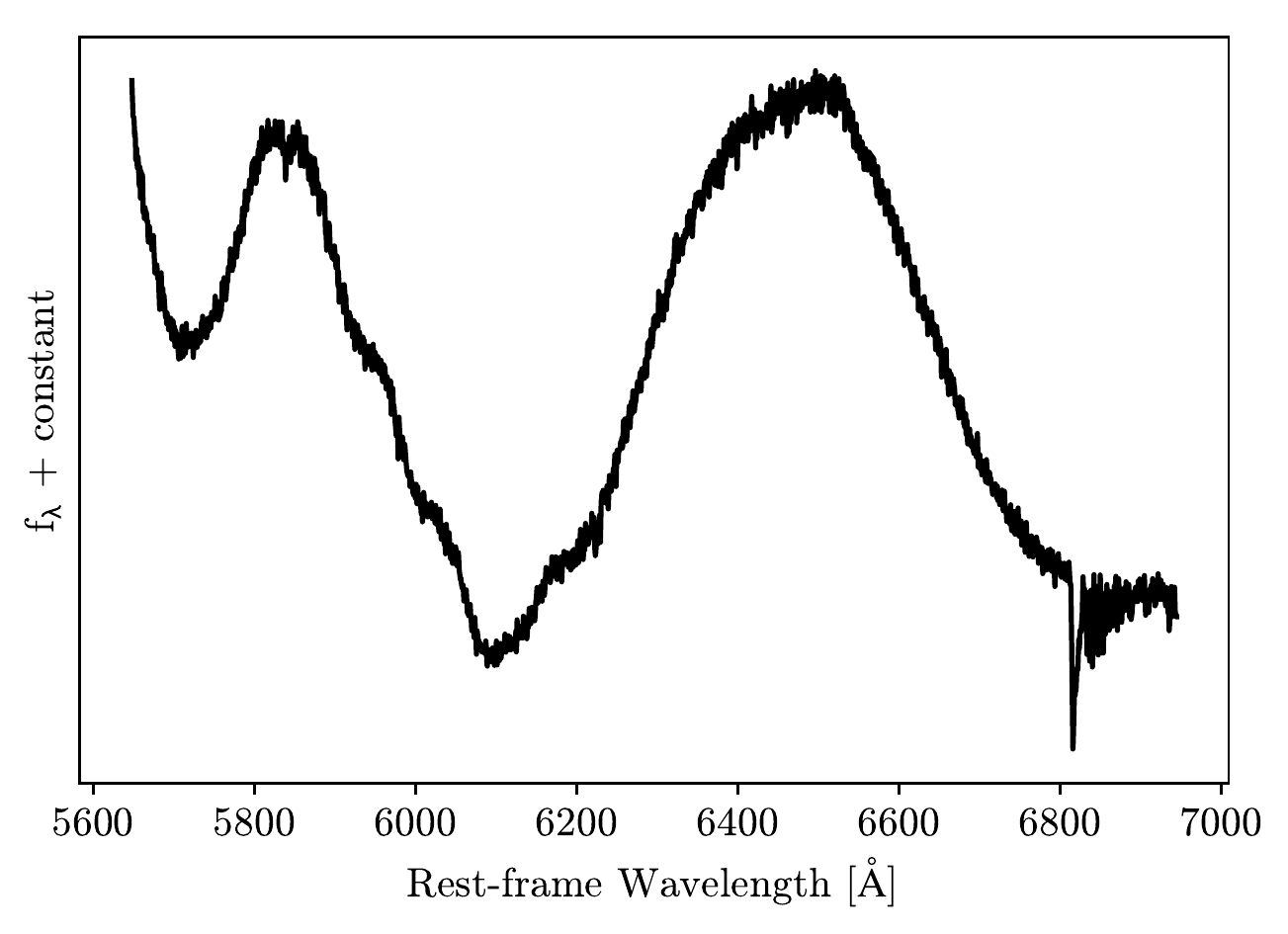}
    \caption{A medium-resolution spectrum of SN\,2019ein, obtained with the MMT spectrograph at +18 days with respect to \textit{B}-band maximum light, centered on the Si II 6355 \AA{} absorption feature. Wavelengths have been shifted to the rest frame.}
    \label{fig:medres}
\end{figure}

Using the MMT Observatory, we obtained a medium-resolution (R $\approx$ 3900) spectrum centered on the Si II 6355 \AA{} absorption feature at +18 days with respect to \textit{B}-band maximum (Figure \ref{fig:medres}). At this phase, the feature takes on an unusual asymmetric appearance. In particular, there appear to be multiple overlapping absorption troughs, each with a different line strength and Doppler shift. This may be caused by significant Si II mixing at this epoch, in which different distributions of Si II are moving at different velocities. This possibility is explored further in Section \ref{sec:discussion}.

\begin{figure}
    \centering
    \includegraphics[scale=0.6]{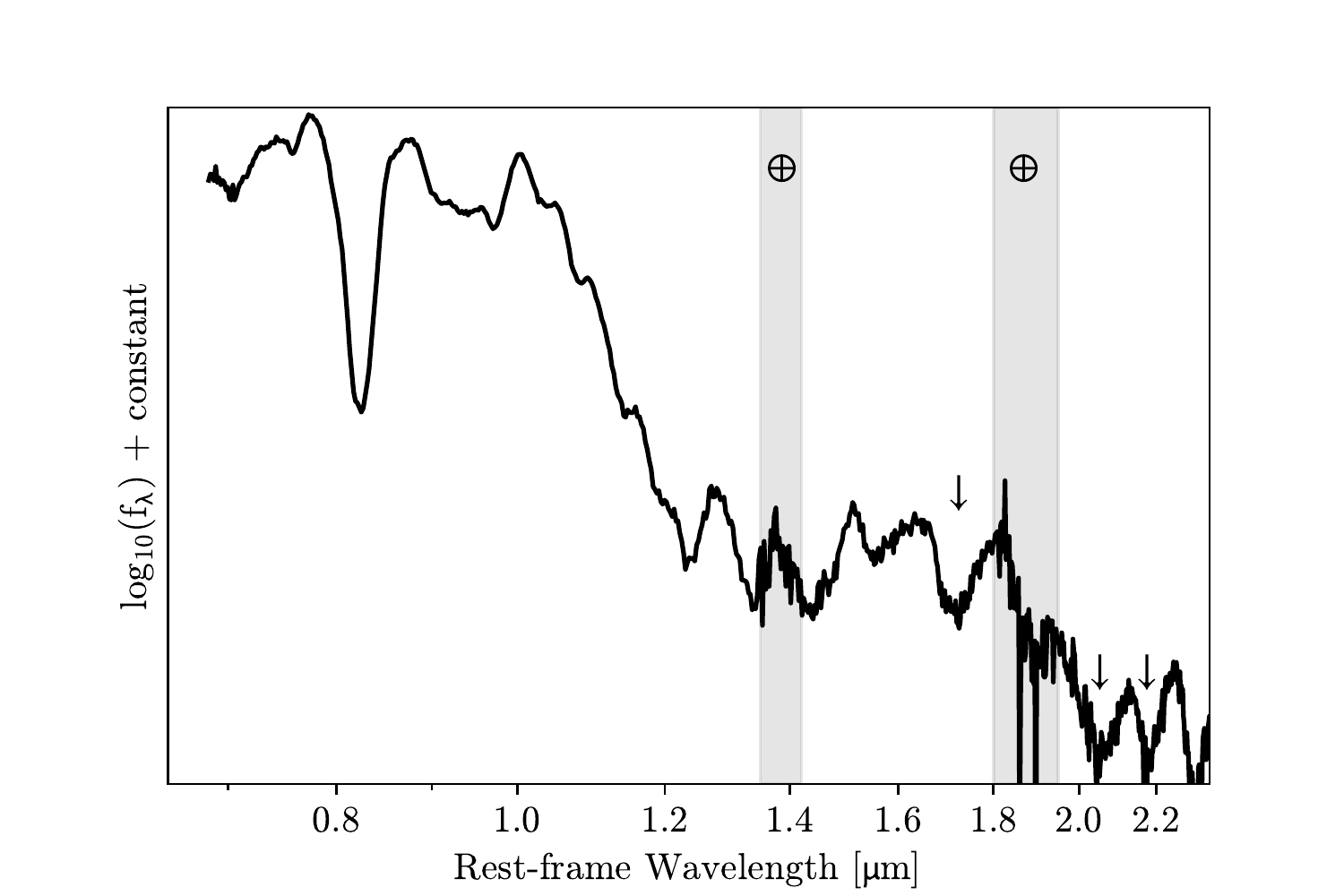}
    \caption{An NIR spectrum of SN\,2019ein, obtained with the SpeX spectrograph via low-resolution PRISM mode on the NASA Infrared Telescope Facility at +32 days with respect to \textit{B}-band maximum light. Wavelengths have been shifted to the rest frame and fluxes have been plotted in logarithmic units. The gray shaded regions denote wavelengths with strong telluric features, while black arrows denote possible IGE absorption features.}
    \label{fig:nir}
\end{figure}

By approximately three weeks after maximum light, the Si II feature begins to blend with iron-group element (IGE) lines that dominate the spectrum. These IGE features, marked with black arrows, are most easily seen in the NIR spectrum obtained 32 days after \textit{B}-band maximum, shown in Figure \ref{fig:nir}. Line blanketing from IGEs are seen in between the two telluric regions and at wavelengths greater than 2.0 $\mu$m. At this later phase, most C I and intermediate-mass element (IME) lines, usually seen around maximum light (e.g. \citealt{Hsiao2013,Hsiao2015}), have disappeared from the NIR spectrum.

\subsection{Branch Classification} \label{subsec:branch}

\begin{figure*}
    \centering
    \includegraphics[scale=0.8]{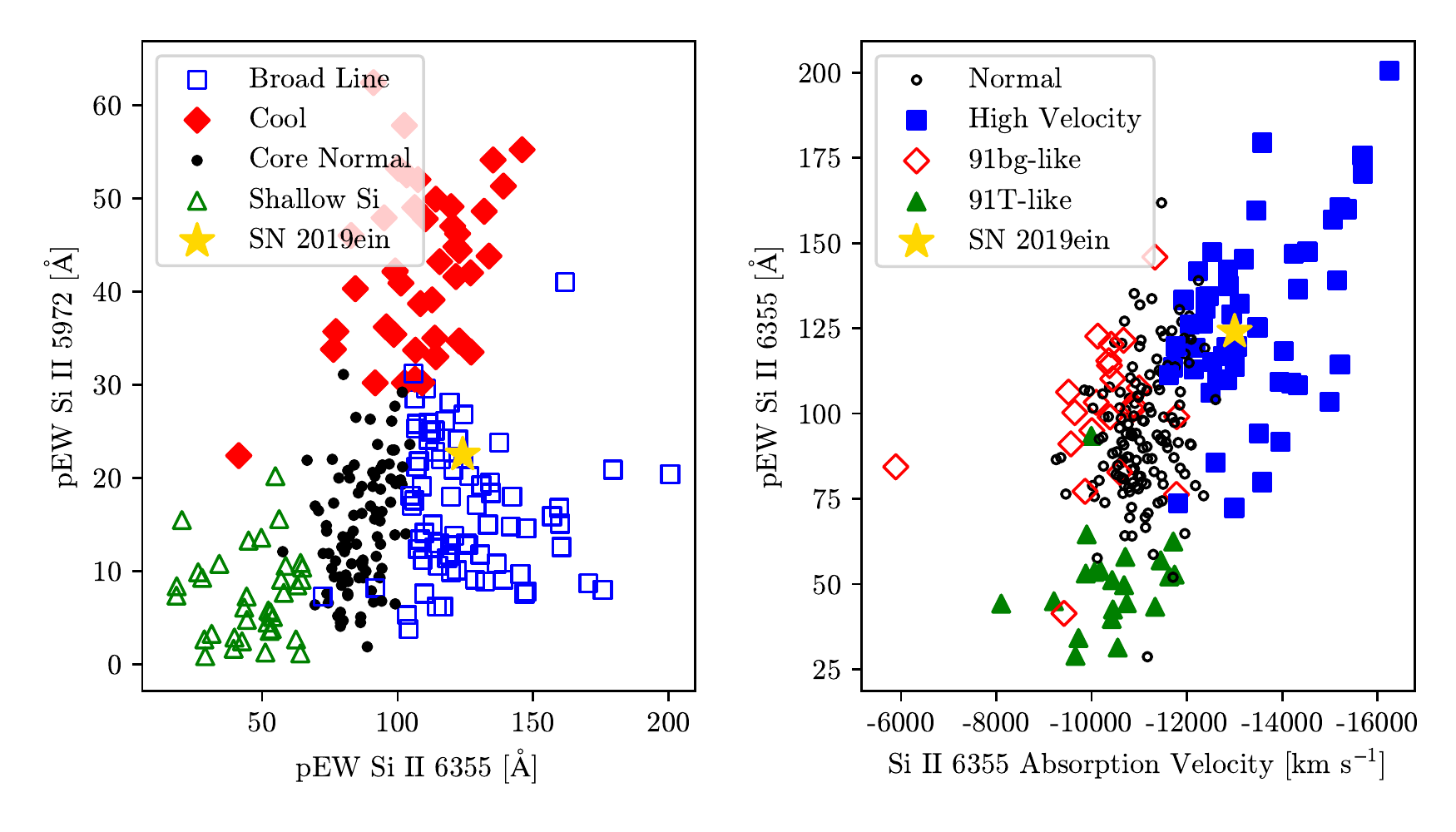}
    \caption{Left: pEW of Si II 5972 \AA{} plotted against the pEW of Si II 6355 \AA{} at \textit{B}-band maximum light, according to \citet{Branch2006}. Different Branch classifications are given by different colored symbols. SN\,2019ein is shown with a gold star. Right: pEW of Si II 6355 \AA{} versus Si II 6355 \AA{} absorption velocity at \textit{B}-band maximum light. Here different symbols correspond to different spectral subclasses of SNe Ia. Sample data are obtained from \citet{Blondin2012}.}
    \label{fig:branch}
\end{figure*}

\citet{Branch2006} showed that the ratio of the pseudo-equivalent width (pEW) of the Si II 6355 \AA{} absorption line to that of the Si II 5972 \AA{} line can be used as a spectroscopic classification of SNe Ia. Here we classify SN\,2019ein in the same way. We measure the pEWs with the following procedure: first, the spectrum is smoothed with a Savitzky-Golay filter to reduce the effects of noise. Next, the absorption feature of interest is defined and maxima blueward and redward of the absorption minimum along the continuum are found. We define the pseudo-continuum as simply the linear curve connecting the two maxima, so long as the curve does not intersect the spectral feature. Finally, the pEW is calculated using the formula (e.g. \citealt{Garavini})
\begin{equation}
    \text{pEW} = \sum_{i=0}^{N-1}\Delta \lambda_i \Bigg(\frac{f_c(\lambda_i) - f(\lambda_i)}{f_c(\lambda_i)}\Bigg)
\end{equation}
, where $f(\lambda_i)$ is the measured flux, $f_c(\lambda_i)$ is the flux of the pseudo-continuum, and $\Delta \lambda_i = \lambda_{i+1} - \lambda_{i}$ is the size of the wavelength bin at each wavelength interval $\lambda_{i}$. 

At maximum light, we find that the pEW of Si II 6355 is 125 $\pm$ 2.1 \AA{} and the pEW of Si II 5972 is 22.5 $\pm$ 2.8 \AA{}. The corresponding Branch diagram is plotted in Figure \ref{fig:branch}. Compared to the sample from \citet{Blondin2012}, SN\,2019ein falls within the broad-line (BL) region of parameter space. This classification agrees with that presented in \citet{Kawabata}. The right side of Figure \ref{fig:branch} shows the pEW of Si II 6355 \AA{} versus the velocity of the Si II 6355 \AA{} absorption feature at maximum light, labeled by spectroscopic subtype. Here SN\,2019ein lies within the population of HV SNe.  \citet{Blondin2012} find a correlation between BL SNe and HV SNe, according to the \citet{Wang2009} classification scheme. 

\subsection{Absorption Velocities}

SN\,2019ein shows some of the highest expansion velocities of any SNe Ia in its early-time spectra. Velocities were calculated following the method outlined in \citet{Childress}: first we select the absorption line of interest and define a pseudo-continuum by fitting a linear curve to the continuum maxima on both sides of the absorption trough. We normalize the flux with respect to this pseudo-continuum before fitting a Gaussian to the normalized absorption line. The minimum of the Gaussian is taken to be the Doppler-shifted observed wavelength, and the expansion velocity is calculated by comparing this measured absorption minimum to the known rest value of the line. 

 \begin{figure}
    \centering
    \includegraphics[scale=0.75]{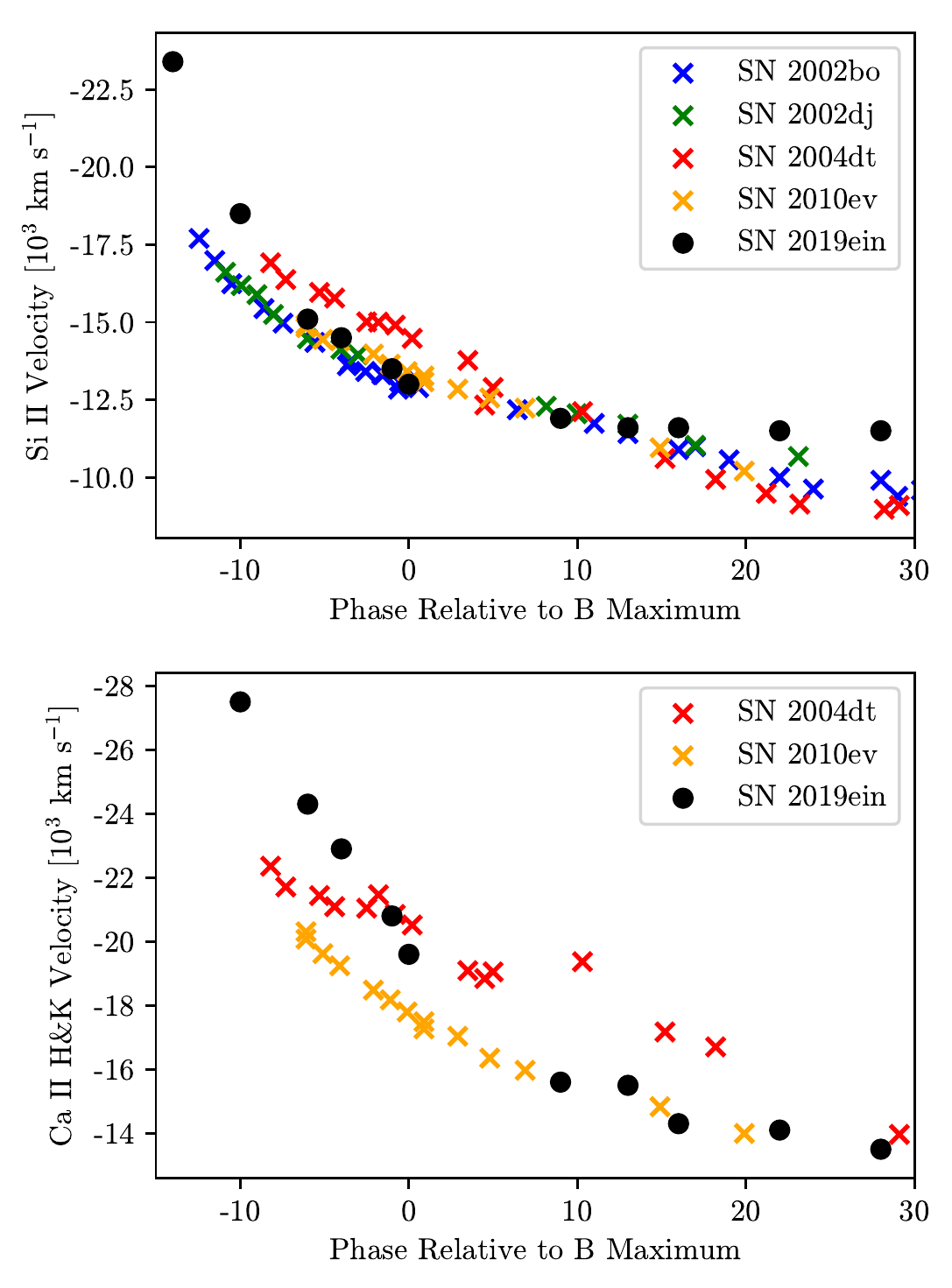}
    \caption{The Si II 6355 \AA{} (top) and Ca II H\&K (bottom) absorption velocity evolution from -14 days to 30 days for SN\,2019ein, compared to a sample of other HV SNe (from \citealt{Gutierrez}).}
    \label{fig:gvels}
\end{figure}

\begin{figure}
    \centering
    \includegraphics[scale=0.62]{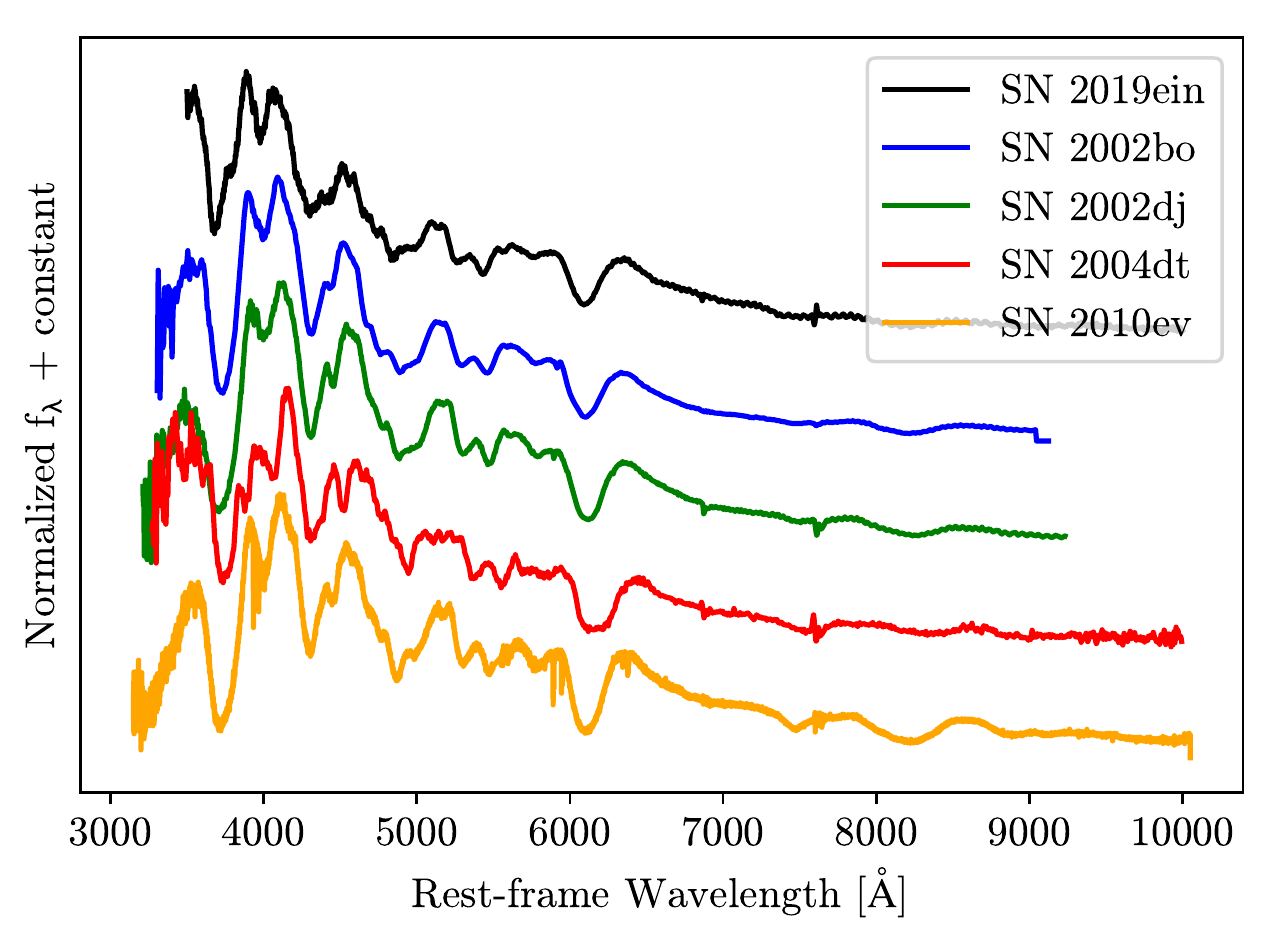}
    \caption{Spectra of SN\,2019ein and four other HV SNe at -4 days with respect to \textit{B}-band maximum light (from \citealt{Gutierrez} and \citealt{Altavilla}).}
    \label{fig:gcomp}
\end{figure}

Figure \ref{fig:gvels} compares the Si II 6355 \AA{} and Ca II H\&K absorption velocity evolution of SN\,2019ein to several other HV SNe Ia from \citet{Gutierrez}. These objects all show similar spectral features (Figure \ref{fig:gcomp}), including strong Ca II NIR absorption at early times, broad Si II 6355 \AA{} absorption at maximum light, and HV Si II and Ca II before maximum. We do not report a Ca II H\&K velocity -14 days with respect to \textit{B}-band maximum light because at this epoch, no clear absorption minimum is identified within the wavelength range of our spectrograph. 

The velocity evolution of all lines is rapid. The first epoch, corresponding to 14 days before maximum light and at most 3 days after explosion, shows the highest Si II velocity in this sample. By maximum light, the ejecta velocity remains high, yet falls within the range of the other HV SNe. After maximum light, we measure a velocity gradient to the Si II velocity following the example of \citet{Blondin2012}, who found that measuring the change in Si II velocity between maximum light and $10 \pm 2$ days after maximum gives the most consistent result. Using this method, we calculate a Si II velocity gradient $\dot{v} = 122 \pm 25$ km s$^{-1}$ day$^{-1}$, placing SN\,2019ein in the HVG class \citep{Benetti2005}. 

\subsection{Comparison to a Delayed-detonation Explosion Model} \label{subsec:explosion}
 
 \begin{figure}
    \centering
    \includegraphics[scale=0.57]{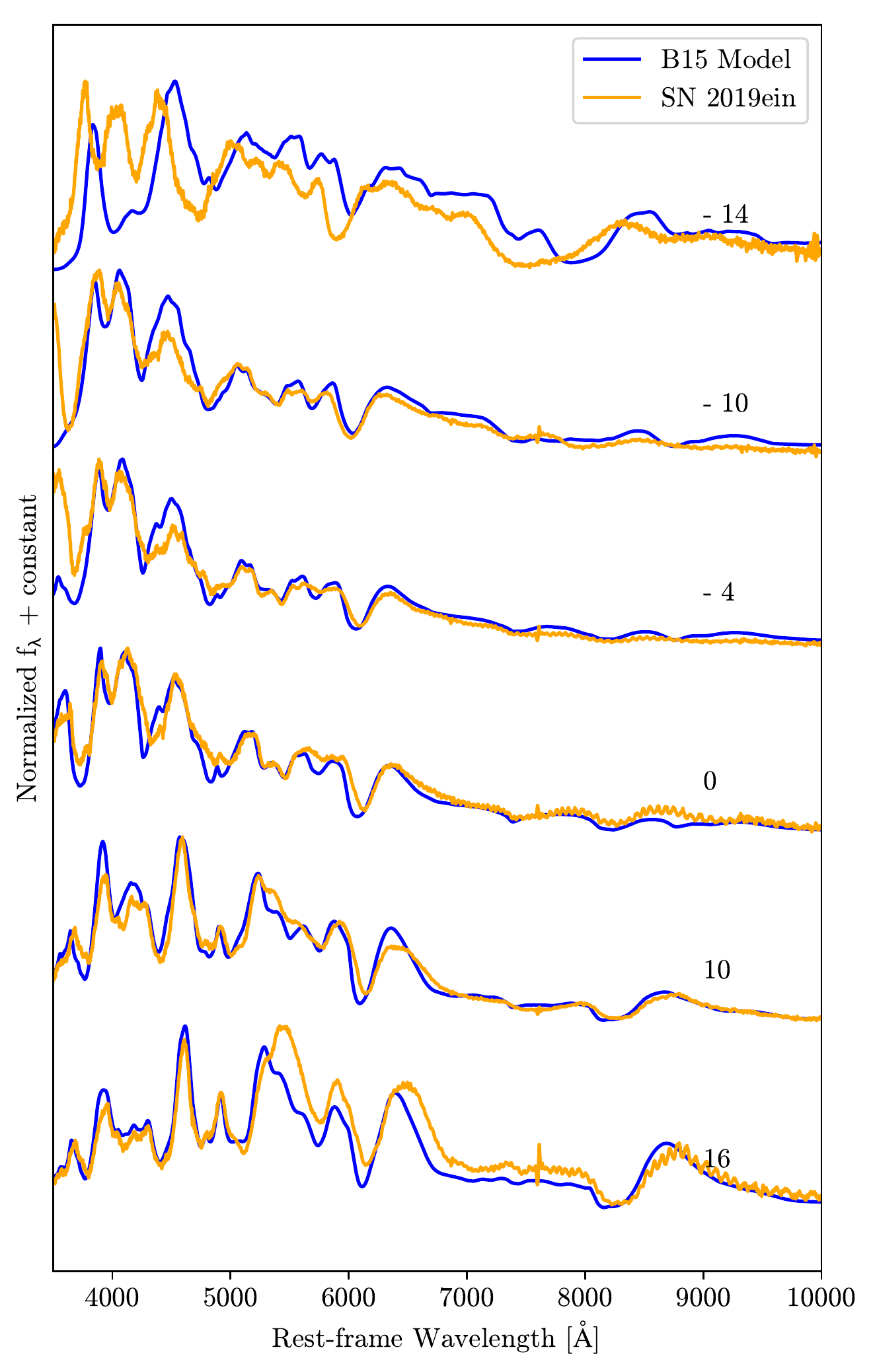}
    \caption{Spectra of SN\,2019ein (orange) compared with the B15 synthetic spectra (blue; \citealt{Blondin2015}). Shown at the top right of every spectrum is the corresponding phase with respect to \textit{B}-band maximum light.}
    \label{fig:synthspectra}
\end{figure}

\citet{Dessart2014} found that delayed-detonation explosions best model BL HVG SNe Ia. Additionally, \citet{Kawabata} found that the observed properties of SN\,2019ein match those seen in the delayed-detonation models of \citet{Iwamoto}. Figure \ref{fig:synthspectra} compares the spectra of SN\,2019ein at various phases to delayed-detonation model spectra produced by \citet{Blondin2015}. The B15 model simulates the spherically symmetric delayed-detonation of a Chandrasekhar-mass WD, imposed with radial mixing to match abundance stratifications observed in SNe ejecta, particularly those of IMEs and IGEs. Synthetic spectra from explosion to nearly 100 days after maximum light are produced.

Beginning 10 days before \textit{B}-band maximum, the synthetic spectra match the strengths and velocities of most of the absorption features, including the Si II 6355 \AA{} and Ca II NIR and H\&K troughs. However, at our earliest epoch of -14 days, the spectrum of SN\,2019ein significantly deviates from the B15 model spectrum at the same phase. The model fails to reproduce the extremely high absorption velocities, the broad mix of O I and Ca II NIR absorption, and the overall blueshift of the emission features with respect to the rest frame of the galaxy. The authors found similar discrepancies when they compared the earliest model spectra to the early-time spectra of SN\,2002bo, and suggested that this may be due to underestimated outward mixing or a more complicated explosion than their one-dimensional, spherically symmetric model. Observational evidence for this enhancement of IMEs in the outer layers of the ejecta, possibly due to an extended burning front or significant mixing, was also found for SN\,2002bo \citep{Benetti2004} and the HV SN\,2004dt \citep{Altavilla}.

In order to investigate the cause of this discrepancy, we now explore possible sources of the HV ejecta in SN\,2019ein, including interaction with a circumstellar shell of material from the progenitor system or mixing and optical depth effects in the ejecta.

\section{Progenitor Constraints from Radio Observations} \label{sec:radio}
\begin{figure*}
    \subfigure[]{\includegraphics[width=0.5\textwidth]{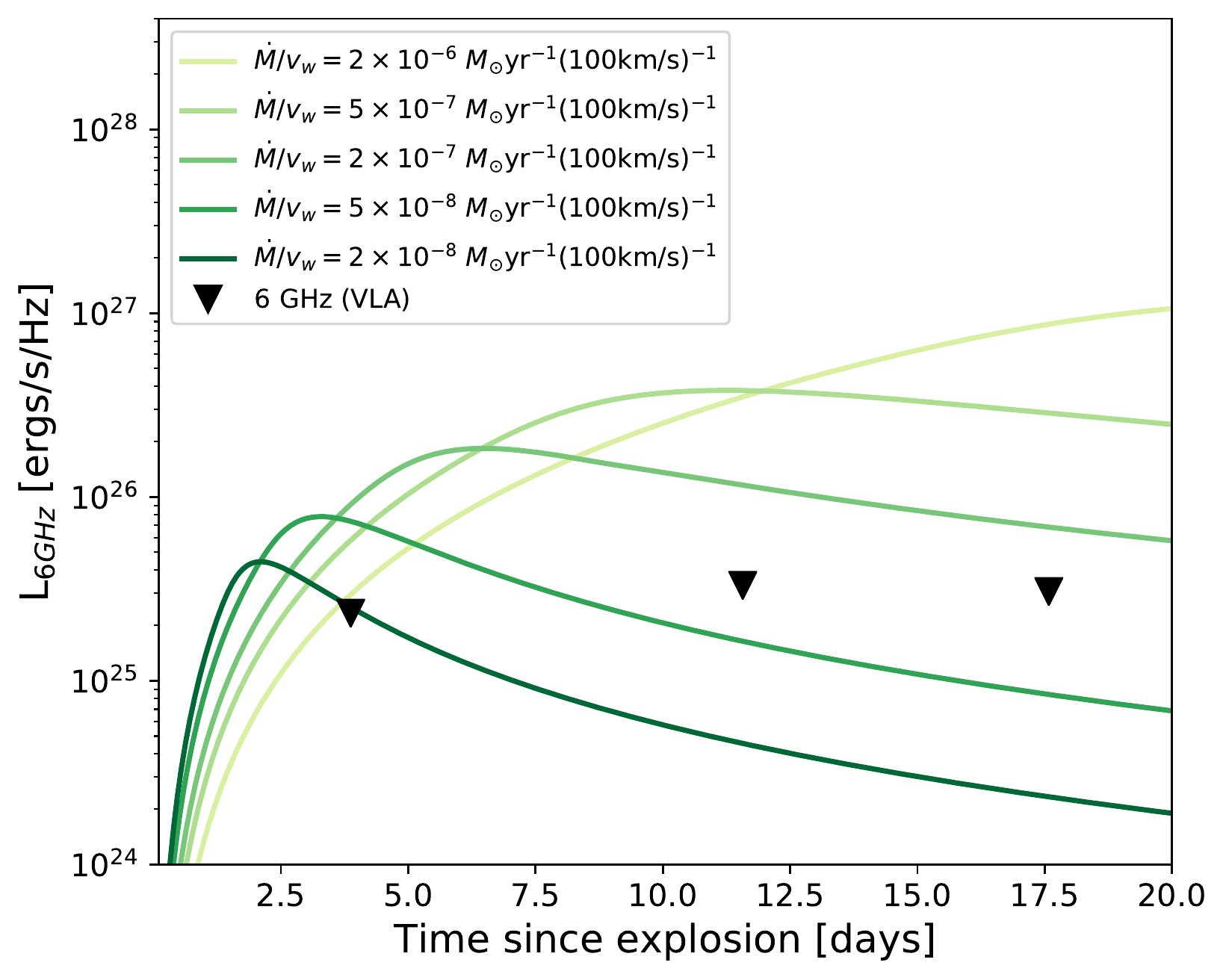}\label{fig:r2lc}}
    \subfigure[]{\includegraphics[width=0.48\textwidth]{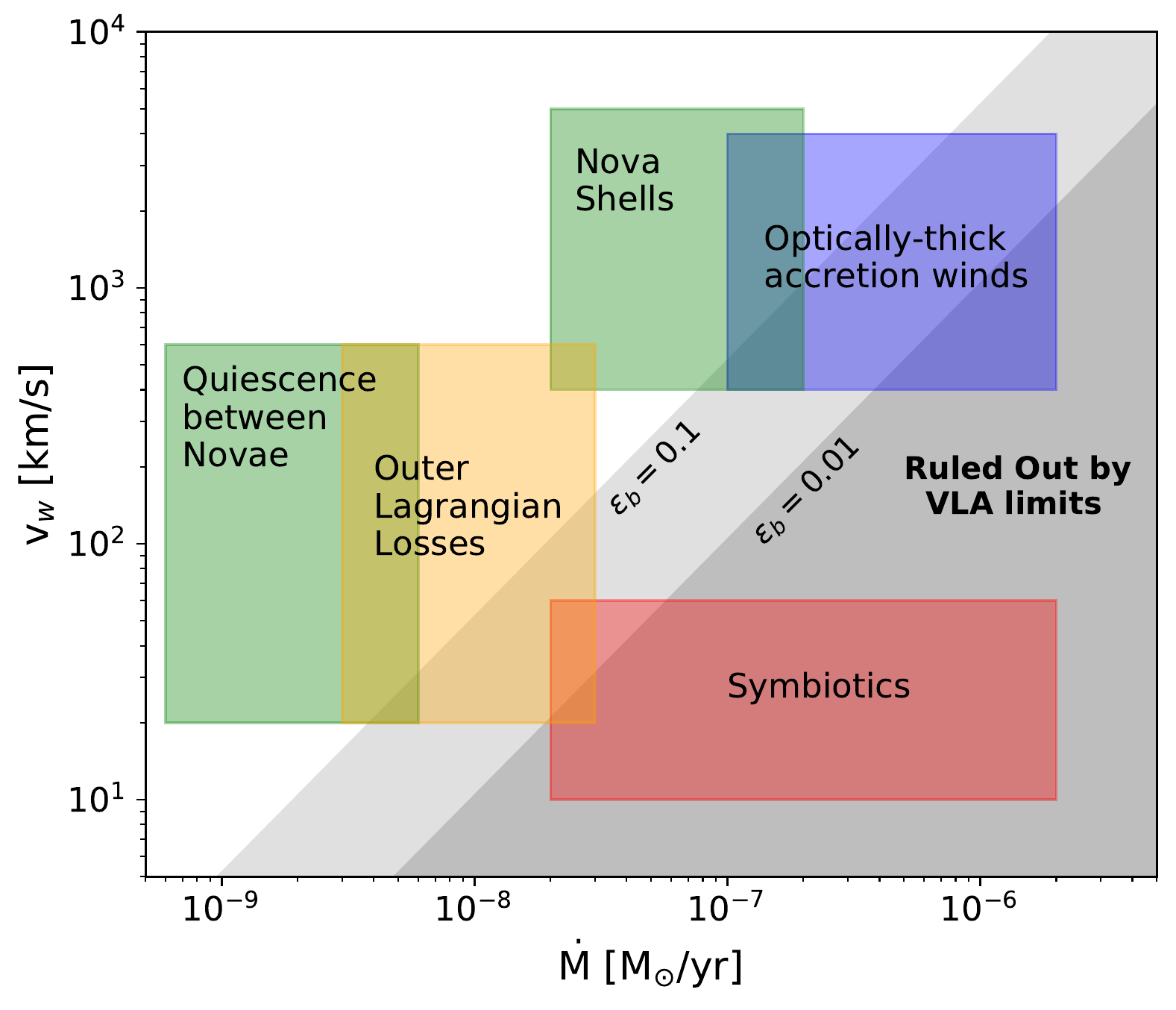}\label{fig:r2models}}
    \caption{(a) 6 GHz radio light curves from the $r^{-2}$ wind model (Section \ref{sec:r2}) for different ratios of constant mass-loss rates ($\dot{M}$) and wind velocities ($v_w$). The 3$\sigma$ 6GHz VLA upper limits are shown as black triangles. (b) The parameter space of $\dot{M}-v_w$. The colored regions show approximate parameter spaces expected for different single-degenerate progenitor models as defined in Figure 3 of \citet{Chomiuk2012}. The light and dark gray regions represent the parameter space of the $r^{-2}$ model that is ruled out by our VLA upper limits. These regions are defined by $\dot{M}/v_w > 1.9 \times 10^{-10}$ M$_{\odot}$ yr$^{-1}$ (km s$^{-1}$)$^{-1}$ assuming $\epsilon_b = 0.1$, and $\dot{M}/v_w > 9.5 \times 10^{-10}$ M$_{\odot}$ yr$^{-1}$ (km s$^{-1}$)$^{-1}$ assuming $\epsilon_b = 0.01$ (see Section \ref{sec:r2} for details)}
\end{figure*}

Radio emission is a sensitive probe of the progenitor environment (which we will refer to as circumstellar medium, or CSM). The CSM is modified by mass loss from the progenitor in the pre-SN stage, and interaction of the SN ejecta with this CSM accelerates electrons to relativistic energies and amplifies the ambient magnetic field, producing synchrotron radio emission \citep{Chevalier1982, Chevalier1984, Chevalier1998}. Simple models of radio emission have provided constraints on the CSM environment and progenitor properties for both core-collapse \citep[e.g.][]{Ryder2004, Chevalier2006, Soderberg2006, Weiler2007, Salas2013} and SNe Ia \citep{Panagia2006, Chomiuk2016}. Radio emission is yet to be detected from a SN Ia, but nondetections have provided stringent constraints on progenitor scenarios \citep{Chomiuk2016}, particularly for nearby events such as SN\,2011fe \citep{Horesh2012, Chomiuk2012} and SN\,2014J \citep{Torres2014}. We can similarly interpret possible progenitor scenarios of SN\,2019ein by comparing our VLA observations with models of radio emission from circumstellar interaction. 

\subsection{Wind Model ($\propto r^{-2}$)} \label{sec:r2}
\begin{figure}
    \centering
    \includegraphics[width=\columnwidth]{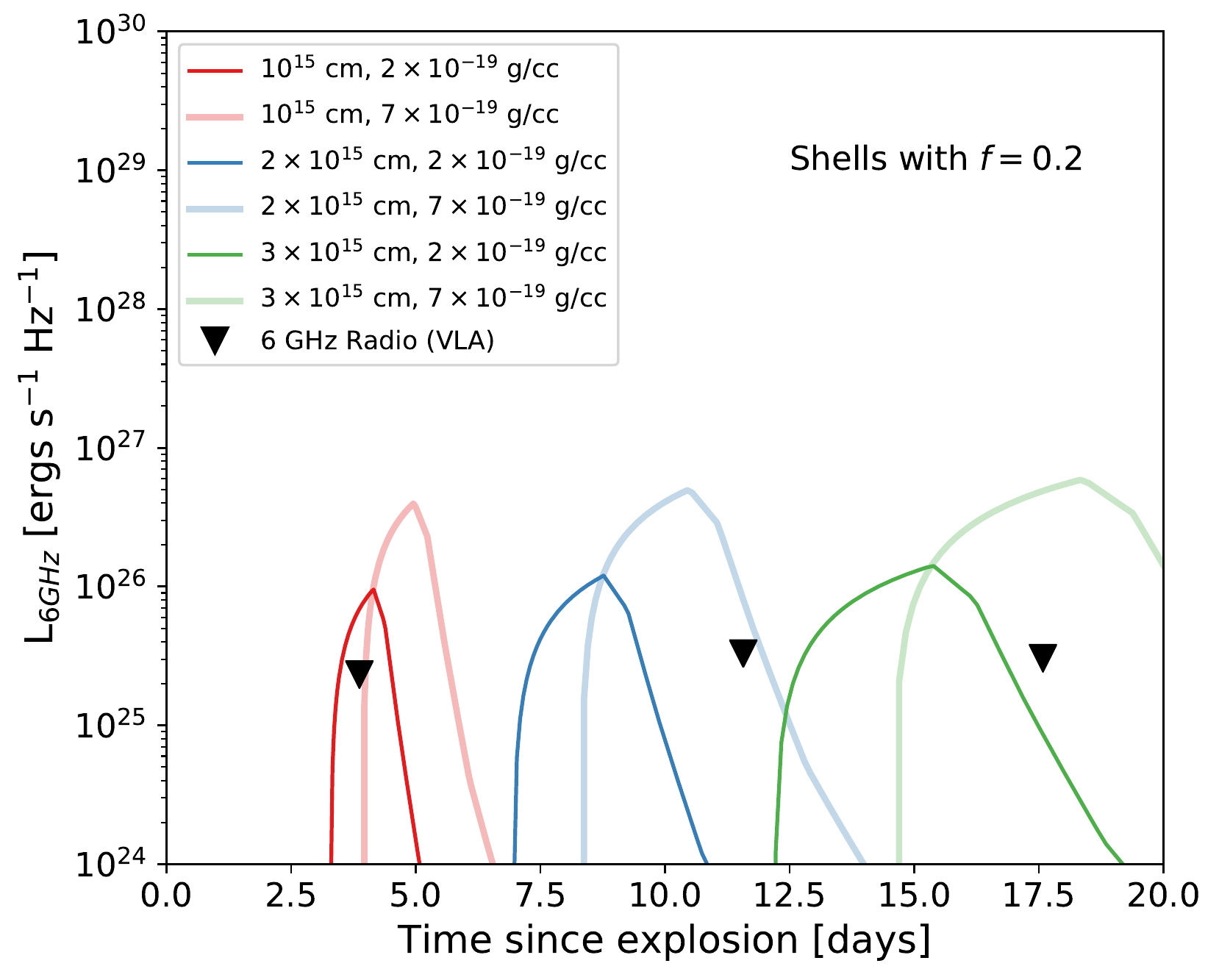}
    \caption{Light curves from the shell model of H16 for a shell of fractional width, $f=0.2$. The colors are for different shell radii, and the different shades are for different densities. The black triangles show the 3$\sigma$ upper limits from our VLA observations.}
    \label{fig:h16lc}
\end{figure}
For single-degenerate progenitors, a fraction of the  mass, transferred via accretion from a nondegenerate companion, is expected to be lost in the form of a wind. \cite{Chevalier1982} created a simple parametric model of such a wind, characterized by a constant mass-loss rate ($\dot{M}$) and wind velocity ($v_w$), which leads to a CSM whose density ($\rho$) varies with radius ($r$) as 
\begin{equation}
\rho = \frac{1}{4 \pi r^2}\left( \frac{\dot{M}}{v_w}\right)
\end{equation}
The synchrotron radio light curve from a shock propagating through such a CSM is described in \cite{Chevalier1982} and \cite{Chevalier1998}. In this work, we follow the formalism of \cite{Chomiuk2016} (hereafter C16), who adopted the self-similar solutions of \cite{Chevalier1982} for radio observations of SNe Ia. We assume a Chandrasekhar-mass WD progenitor that exploded with 10$^{51}$ erg of kinetic energy, consistent with our optical observations, and a steep outer ejecta profile of $\rho_{ej} \sim v_{ej}^{-10}$ interacting with the above CSM. Electrons are accelerated to a power-law spectrum ($\sim E^{-p}$, with $p=3$). The average fraction of the shock energy shared by the cosmic-ray electrons and the amplified magnetic field in the shock vicinity is parameterized as $\epsilon_e$ and $\epsilon_b$ respectively. As in \citet{Chomiuk2012} and \citet{Chomiuk2016}, we set $\epsilon_e=0.1$ and $\epsilon_b = [0.1, 0.01]$, consistent with values expected in Type Ib/c SNe \citep{Chevalier2006, Sironi2011, Soderberg2012}. 

The light-curve models for different values of the free parameters $\dot{M}$ and $v_w$ are shown in Figure \ref{fig:r2lc}. The rising part of the light curves corresponds to the regime where the ejecta are still optically thick to synchrotron self-absorption at 5 GHz. When the ejecta are optically thin, the light curve declines. Higher ratios of $\dot{M}/v_w$ correspond to denser outflows, which leads to brighter light curves and a delayed transition to the optically thin stage, which explains why the peaks are shifted to later epochs. 

Figure \ref{fig:r2models} shows our constraints on the $\dot{M}/v_w$ parameter space from the VLA upper limits in Section \ref{sec:radioobs}. We are able to rule out the parameter space for $\dot{M}/v_w > 1.9 \times 10^{-10}$ M$_{\odot}$ yr$^{-1}$ (km s$^{-1}$)$^{-1}$ for $\epsilon_e = \epsilon_b = 0.1$. For $\epsilon_e = 0.1$ and $\epsilon_b = 0.01$, we find that $\dot{M}/v_w > 9.5 \times 10^{-10}$ M$_{\odot}$ yr$^{-1}$ (km s$^{-1}$)$^{-1}$.

The above constraints on $\dot{M}/v_w$ can provide some insight into possible single-degenerate progenitor models for SN\,2019ein by comparing with typical values of $\dot{M}$ and $v_w$ expected in these models as compiled in \cite{Chomiuk2012}. Our observations are sensitive enough to rule out symbiotic progenitors, i.e. a WD that accretes from the wind of a giant companion, which is generally characterized by $\dot{M}>10^{-8}$ M$_{\odot}$ yr$^{-1}$ and $v_w \approx 30$ km s$^{-1}$ \citep{Seaquist1990, Chen2011, Patal2011}. A symbiotic channel was also deemed unlikely for the nearest events SN\,2011fe and SN\,2014J, and was found to contribute no more than $16\%$ of a sample of 85 SNe Ia with available radio observations studied by \citep{Chomiuk2016}. For $\epsilon_e = 0.1$ and $\epsilon_b=0.01$, our increased upper limit of $\dot{M}/v_w > 9.5 \times 10^{-10}$ M$_{\odot}$ yr$^{-1}$ (km s$^{-1}$)$^{-1}$ still excludes the majority of symbiotic progenitors observed in the Galaxy \citep{Seaquist1993, Chomiuk2016}. 

White dwarfs can also be in single-degenerate systems with a main-sequence or a slightly evolved companion undergoing mass transfer via Roche-lobe overflow. For mass accretion rates $\gtrsim 3 \times 10^{-7}$ M$_{\odot}$ yr$^{-1}$, steady nuclear burning occurs on the surface of the WD, and about $\sim 1\%$ of the mass is lost from the outer Lagrangian point with velocities of about a few 100 km s$^{-1}$ \citep{Shen2007}. The expected $\dot{M}/v_w$ in such a scenario falls within our VLA limits, and therefore such a progenitor channel cannot be ruled out for SN\,2019ein from our radio observations alone. With increasing accretion rate, however, the nuclear burning shell will drive fast optically thick winds with $v_w \approx$ few $\times 1000$ km s$^{-1}$ \citep{Hachisu1999}, and some part of this parameter space is ruled out by our VLA upper limits. For accretion rates $\approx (1-3) \times 10^{-7}$ M$_{\odot}$ yr$^{-1}$, the steady burning will be interrupted by recurrent nova flashes. Novae with short recurrence time will likely create a series of dense shells with which the SN shock will interact, and such shells typically have values of $\dot{M}$ and $v_w$ as shown in Figure \ref{fig:r2models}. For longer recurrence times, the SN shock is more likely to interact with CSM created with a steady wind with $\dot{M} \approx 10^{-9}-10^{-8}$ M$_{\odot}$ yr$^{-1}$ in between distant novae shells \citep{WoodVasey2006}. Both these cases are allowed within our upper limits in the context of the $r^{-2}$ model, but we will also analyze the presence of nova shells with a more appropriate shell-interaction model in Section \ref{sec:h16}.

We note here the importance of radio observations taken soon after explosion or discovery for SNe Ia. The first observation, which was triggered $<2$ days of discovery and $\lesssim 4$ days of explosion, provided a constraint that is almost a factor of five deepert on $\dot{M}/v_w$ than the observation a week later (Figure \ref{fig:r2lc}). This is because lower $\dot{M}/v_w$ shifts the peak of the radio light curve to earlier times, as seen in Figure \ref{fig:r2lc}. The prompt observation resulted in more stringent constraints on Type Ia progenitor models involving symbiotic systems and optically thick winds.

\subsection{Shell Interaction Model} \label{sec:h16}
\begin{figure*}
    \centering
    \includegraphics[width=0.8\textwidth]{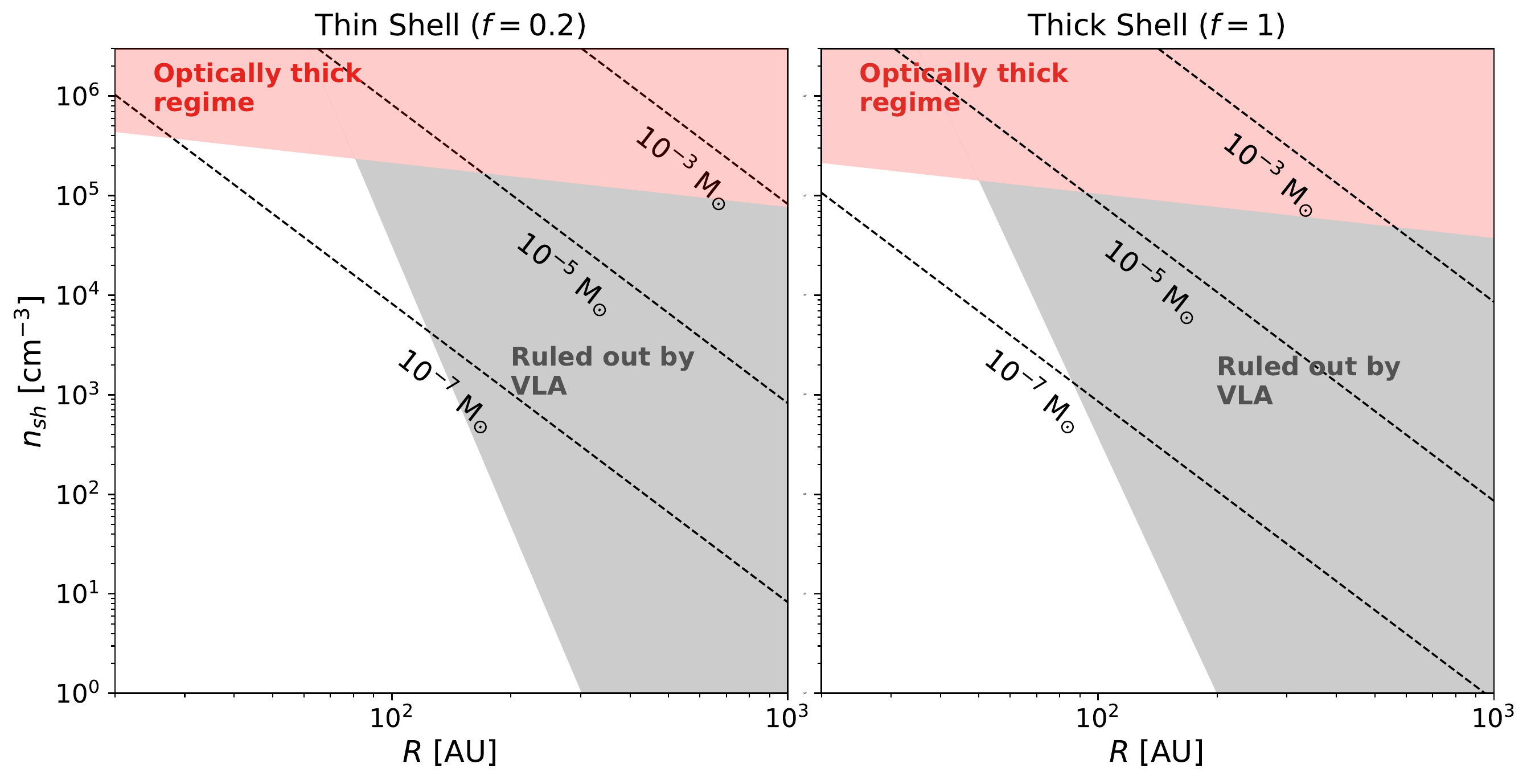}
    \caption{Parameter space of the CSM shell radii $R$ versus shell density $n_h = \rho_{sh}/\mu m_p$, where $\rho_{sh}$ is the density in units of g cm$^{-3}$, $\mu=1.4$, and $m_p = 1.67 \times 10^{-24}$ g. Dashed lines in both panels correspond to lines of constant shell mass $M_{sh} = 4/3 \pi \rho_{sh} R^3 [(1 + f)^3 - 1]$. The left panel corresponds to a thin shell, while the right panel corresponds to a thick shell. The gray shaded region is the parameter space where the H16 light curves are inconsistent with the constraints described in Section \ref{sec:h16}. The red shaded region is where optical depth due to synchrotron self-absorption is $> 1$ and the optically thin ejecta assumption of H16 is no longer valid.}
    \label{fig:h16paramspace}
\end{figure*}
Interaction and acceleration of material in solar-composition CSM shells has been proposed as a way to explain HVFs in Type Ia spectra \citep{Gerardy,Mulligan2018,Mulligan2017}. Such shells can be expected in WD progenitors that undergo nova outbursts. Shells can consist of recently ejected material or of swept-up material from previous outbursts. Shock interaction with such a shell can produce detectable radio emission, and radio light curves for such a CSM created by discrete mass-loss events cannot be appropriately described by a continuous mass-loss model.

We therefore use the models described in \cite{Harris2016}, hereafter H16, for radio emission from a CSM shell interacting with SN ejecta. H16 performed hydrodynamical simulations of a $\rho_{ej} \sim v_{ej}^{-10}$ ejecta profile interacting with a single, solar-metallicity, fully ionized shell defined by an inner radius $R$, fractional width $f = \Delta R/R$, and constant shell density $\rho_{sh}$. Interaction creates a forward shock in the shell and a reverse shock in the ejecta, but the dynamics do not reach self similarity, unlike in the \cite{Chevalier1982} case. The forward shock subsequently accelerates the CSM and sets it in free expansion. 

In the optically thin approximation, the H16 light-curve model can be analytically expressed in terms of $f, \rho_{sh}$ and $R$ (see Eqs. 5-13 in H16). Figure \ref{fig:h16lc} shows example light curves from the H16 model. The light curves are characterized by a rapid brightening at the beginning of the interaction, reaching a peak luminosity when the forward shock reaches the outer edge of the CSM shell, and a steep decline once the shock breaks out.\footnote{The model assumes a vacuum outside the shell region. The decline phase will therefore likely be modified when there is a progenitor wind present beyond the shell (C. E. Harris et al. 2020, in preparation).} For larger $f$, the light curves peak at later times because the shock takes longer to reach the CSM outer edge. For larger $R$, the light curves begin at a later time, and larger $\rho_{sh}$ produces brighter light curves. 

Similar to the analysis in \citet{Cendes2020}, we explore the parameter space of $R$-$\rho_{sh}$ for a given $f$ that produces light curves within our VLA upper limits at the observed epochs. We explore two cases of shells: a thin shell ($f=0.2$) characteristic of shells expected in nova eruptions, and a thick shell ($f = 1$) to show the effects of increasing shell width. Similar to the wind model, the shell models assume a standard Chandrasekhar-mass WD explosion with 10$^{51}$ erg of kinetic energy, and $\epsilon_e = \epsilon_b = 0.1$. We also use an additional constraint: the peak of the light curve must occur before the first epoch (i.e. at 3.87 days after explosion). This is because any shell interaction leading to HV absorption features must have occurred before the first spectral observation \citep[i.e. after the shell has been accelerated by the forward shock,][]{Gerardy}.

Figure \ref{fig:h16paramspace} shows the result of applying the H16 shell models to our VLA observations. For our fiducial model parameters in both the thin and thick shell cases, the VLA limits only allow CSM shells $\lesssim 10^{-6}$ M$_{\odot}$ within radii $<100$ AU. In comparison, CSM masses $\sim 10^{-3}-10^{-2}$ M$_{\odot}$ are generally required to explain HVFs observed in SNe Ia \citep{Gerardy}. We note that this conclusion remains unchanged even when we assume $\epsilon_b = 0.01$ because the shaded region in Figure \ref{fig:h16paramspace} is determined primarily by the condition that the peak of the light curve must occur before the first epoch, as mentioned previously. As explained in H16, the light curve peaks when the shock reaches the outer edge of the shell and is thus mainly a hydrodynamical timescale, which is independent of the parameter $\epsilon_b$ that affects only the radio emission.

A caveat, however, is that the H16 model approximates the radio emission as optically thin, whereas at densities $>10^{5}$ cm$^{-3}$ for the radii explored here, effects such as synchrotron self-absorption, free-free absorption, and radiative cooling of the shock gas will become important. Light-curve models in this optically thick regime would require a formal solution of the radiative transfer equation, which will be explored in an upcoming paper (C. E. Harris et al. 2020, in preparation), and will help provide more accurate constraints on the presence of dense and massive CSM shells.

\begin{figure*}
    \centering
    \includegraphics[scale=0.75]{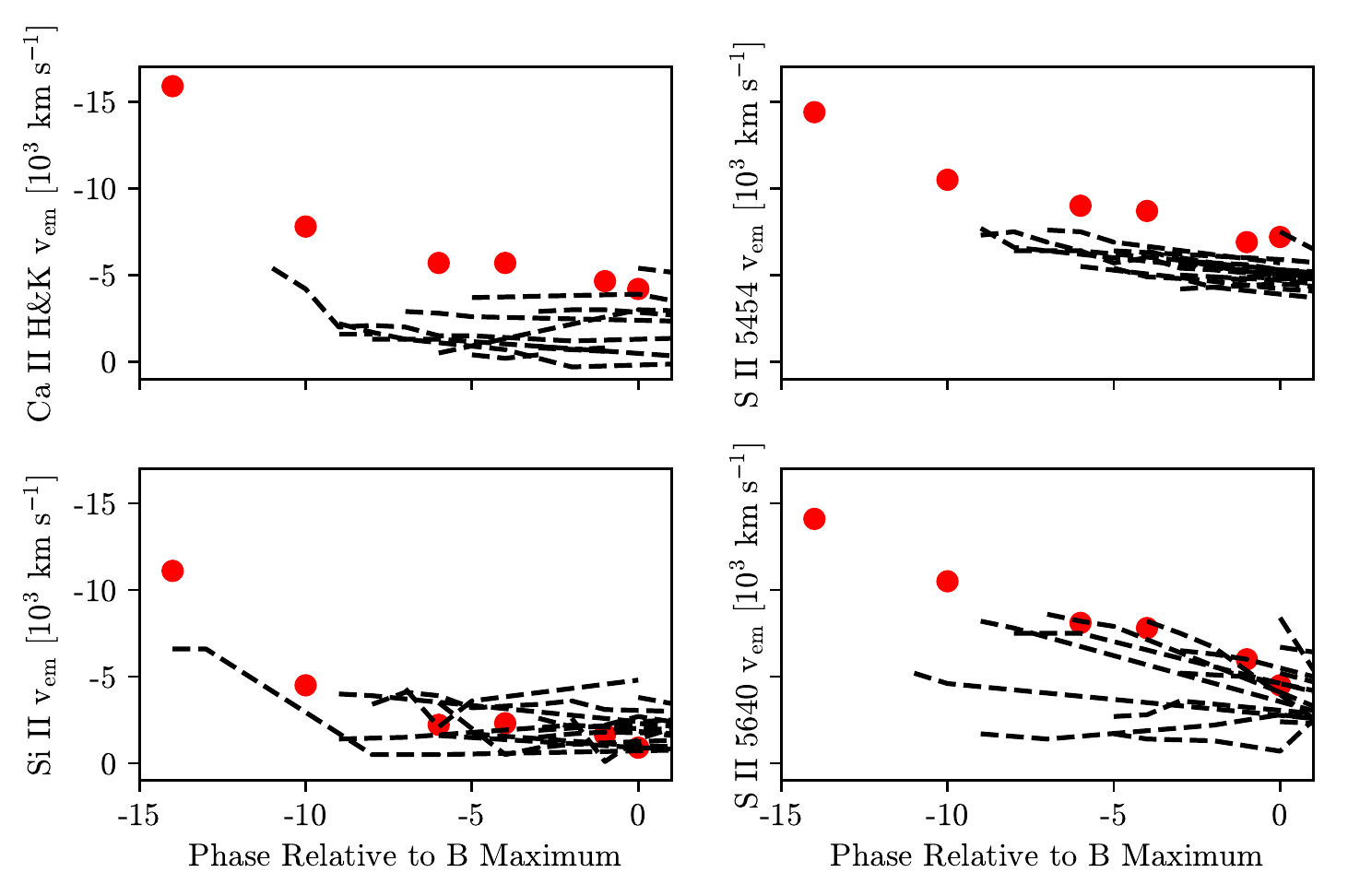}
    \caption{The emission peak Ca II H\&K (top left), Si II 6355 \AA{} (bottom left), S II 5454 \AA{} (top right), and S II 5640 \AA{} (bottom right) velocities of SN\,2019ein, shown in red, compared against the low-redshift sample from \citet{Blondin2006}, shown in dashed black, from -14 days to 0 days with respect to \textit{B}-band maximum light.}
    \label{fig:emvel}
\end{figure*}
 
 \section{Mixing and Optical Depth Effects}\label{sec:discussion}
 
The high emission velocities before maximum light make SN\,2019ein unusual, even among HV SNe Ia. More specifically, although P Cygni emission blueshifts have been theoretically predicted and observed in Type II SNe \citep{Dessart2005} and were discussed by \citet{Blondin2006} in a sample of low-redshift SNe Ia, the emission velocities seen in the spectra of SN\,2019ein are the highest ever measured. Figure \ref{fig:emvel} shows the evolution of the emission velocities for four lines in the spectra of SN\,2019ein compared to the sample from \citet{Blondin2006}. It is clear that at early times, the emission peaks in SN\,2019ein are substantially more blueshifted than in any of the objects in the comparison sample. This extreme behavior is most clearly seen in the plots of the Ca II H\&K and S II emission velocities, where the emission velocities at -14 days with respect to \textit{B}-band maximum are $\approx$ 15,000 km s$^{-1}$. At the same phase, the Si II emission component of the P Cygni profile is blueshifted by $\approx$ 10,000 km s$^{-1}$. Even around maximum light, the velocities of these lines are among the highest ever measured. After maximum, the emission peaks are either no longer resolvable or become distorted due to line overlap, possibly of multiple Doppler-shifted emission features (see Figure \ref{fig:medres}). Here we only present emission velocities up to \textit{B}-band maximum, where we trust our measurements have not been biased.
 
In order to investigate whether specific ejecta compositions or abundance enhancements could cause both the high absorption and emission velocities at early times, we compare \texttt{SYN++} \citep{Thomas} model spectra to our spectrum of SN\,2019ein at -10 days. In particular, we focus on the Si II 6355 \AA{} feature and test whether multiple components of the ejecta, such as a HV component with a velocity above the photospheric velocity (PV), can reproduce the measured Doppler shifts. 

\begin{figure*}
    \centering
    \includegraphics[scale=0.6]{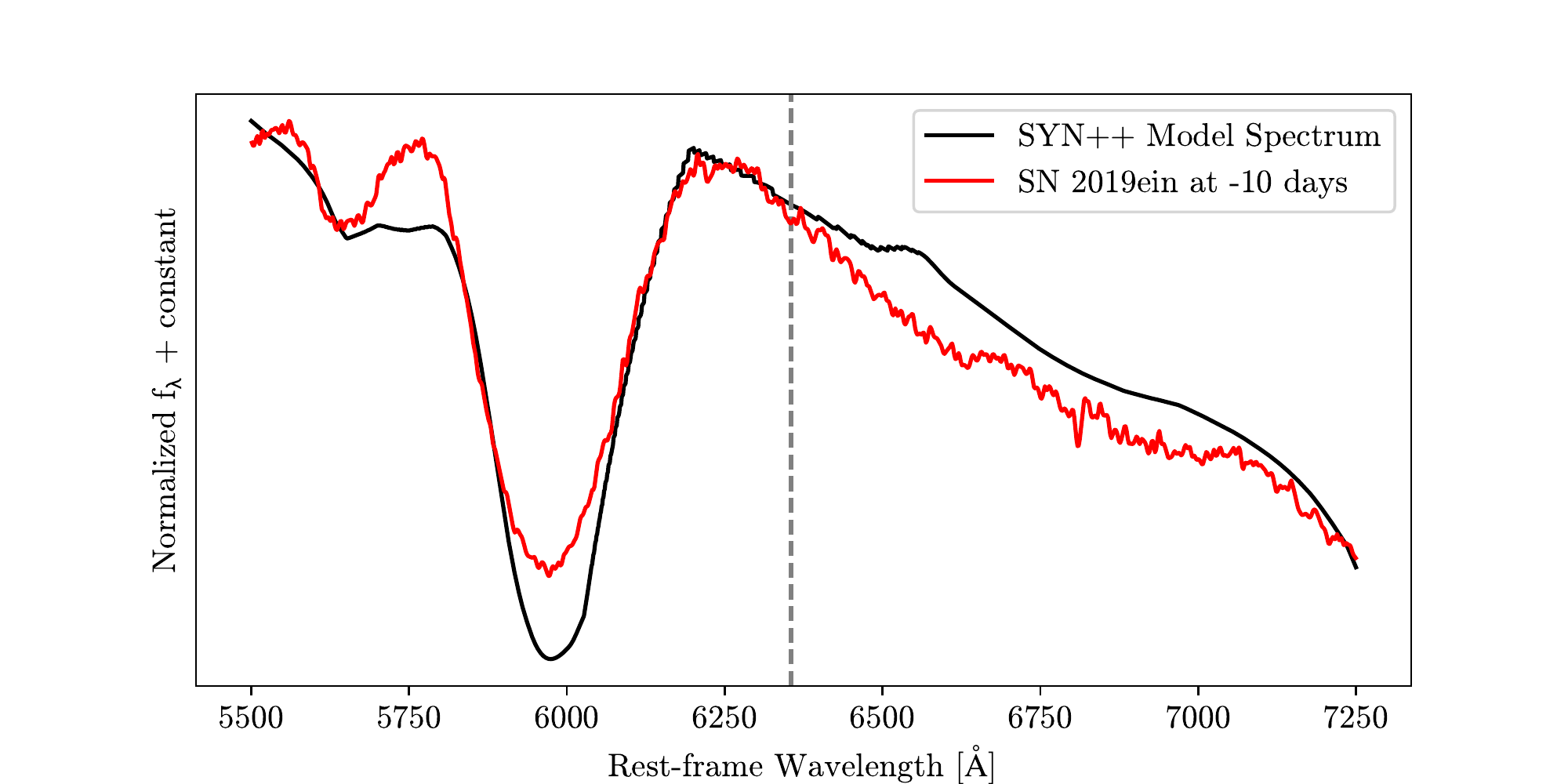}
    \caption{\texttt{SYN++} synthetic spectrum (black) of the Si II 6355 \AA{} feature at -10 days with respect to \textit{B}-band maximum, compared to the spectrum of SN\,2019ein at the same epoch (red). Our synthetic spectrum contains only a HV Si II component offset from the photospheric velocity by 2,500 km s$^{-1}$. The rest wavelength of the Si II emission peak is represented by a dashed vertical line.}
    \label{fig:syn++}
\end{figure*}

Our synthetic spectrum is shown in Figure \ref{fig:syn++} compared to our spectrum of SN\,2019ein at -10 days. We find that an ejecta with only a HV Si II component offset from the PV by several thousand km s$^{-1}$ provides the best fit to our data. This matches the lack of separate HVFs and PVFs, particularly in the Si II and Ca II lines, at early times in our observed spectra, as well as the analysis of our radio observations, which places stringent constraints on the mass of a CSM shell, which has been proposed to produce HVFs. Altogether, this indicates that only a HV Si II component is present in the ejecta of SN\,2019ein. We attempt to explain the existence of this HV component as being due to ejecta mixing from an asymmetric explosion or being caused by optical depth effects in the outer layers of the ejecta. 
 
\subsection{Evidence for Asymmetries}
 
In Section \ref{subsec:explosion} we discuss that objects similar to SN\,2019ein exhibit significant mixing of their IMEs to higher velocities. This may be evidence of an aspherical ejecta distribution due to an asymmetric explosion, in which clumps of IMEs are mixed to higher velocities along the observer's line of sight, producing HV absorption and emission features. Similar clumps are produced in models of off-center delayed-detonation explosions \citep{Seitenzahl}. The connection between mixing and asymmetries has observational support. Polarization measurements show that HVFs in SNe ejecta are more polarized than PVFs, indicating that HV ejecta have more asymmetric distributions \citep{Maund2013,Bulla2016}. Additionally, \citet{Nagao} observed high polarization alongside a blueshift of the H$\alpha$ line during the photospheric phase of the Type II SN\,2017gmr. 

There is evidence for such an aspherical ejecta distribution in the spectra of SN\,2019ein; in the early-time spectra of SN\,2019ein, we see Si II with a higher absorption velocity than C II at -14 days, whereas the opposite relation is true at this phase for the SNe studied by \citet{Parrent2011}. In a spherically symmetric model of a Type Ia explosion, the inner burning regions are surrounded by a shell of unburnt material comprised mostly of C and O. However, an asymmetric explosion with strong outward mixing could force IMEs produced in the nuclear burning to higher velocities. In addition, our medium-resolution spectrum obtained 18 days after \textit{B}-band maximum light reveals multiple overlapping Si II absorption features, with each absorption minimum offset by several thousand km s$^{-1}$. This may be evidence of significant mixing of the Si II ejecta to lower and higher velocities, rather than the stratified shell-like structure proposed in studies of other SNe Ia \citep{Cain2018}. 

Studying the nebular-phase spectroscopy of SNe Ia, \citet{Maeda2010} found a relationship between the Doppler shift of the nebular-phase emission features and the Si II velocity gradient at early times, whereby HVG SNe exhibit redshifted nebular-phase emission lines and LVG SNe show blueshifted nebular emission lines. This correlation is suggested to detail information about the symmetry of the explosion because the nebular-phase emission features trace the deflagration ash in the core of the progenitor WD. In this model, HVG SNe are viewed from the direction opposite to the initial deflagration. 

Finally, \citet{Maund} and \citet{Cikota} found correlations between the Si II line polarization and velocity evolution around maximum, with more polarized SNe belonging to the HV and HVG classes. \cite{Maund} argue that this relationship implies the existence of global asymmetries in the ejecta, which, along with the correlation between velocity evolution and nebular-phase velocity shifts \citep{Maeda2010}, connects early- and late-time velocity behavior to the three-dimensional geometry of the explosion. It is possible that the high absorption and emission velocities seen in the spectra of SN\,2019ein are signatures of IMEs in the ejecta that were outwardly mixed in an off-center explosion.

\subsection{Optical Depth Effects}

Blueshifted emission features have been suggested to be caused by optical depth effects in Type II SNe (e.g. \citealt{Dessart2005,Dessart2011,Anderson}). These features arise from steep density profiles in the expanding ejecta (see Figure 16 in \citealt{Dessart2005}). \citet{Blondin2006} were the first to model these features in SNe Ia. Using the radiative transfer code \texttt{CMFGEN} \citep{Hillier}, the authors found that differences in optical depths of Si II and S II lines resulted in overall blueshifted emission peaks when the flux was integrated over a range of impact parameters. 

In this picture, contours of constant optical depth in the photosphere trace out a variable amount of the emitting ejecta, where the amount of emission from the ejecta above the photosphere depends on the optical depth of the line being considered. In the classical P Cygni profile, the flux from the ejecta moving perpendicular to the observer's line of sight makes up the majority of the emission, resulting in an emission peak centered on the rest wavelength of the line. However, when lines with low optical depth are considered, there is little to no emission from the ejecta at large impact parameters because the density gradient in the outer ejecta layers is steep. Instead, the flux is dominated by ejecta moving toward the observer even if the ejecta is distributed more or less spherically. The result is an overall blueshift of the emission peak, proportional to the ejecta velocity. Because the authors modeled individual lines, the emission blueshifts cannot be the result of line overlap.

As seen in Figure \ref{fig:emvel}, SN\,2019ein has some of the highest emission peak velocities at early times compared to the sample from \citet{Blondin2006}. This extreme behavior can be understood in the context of the above explanation: because blueshifted emission is dominated by flux from the ejecta moving toward the observer, we would expect that the emission velocity is correlated with the absorption velocity. \citet{Blondin2006} found that this trend exists, with the ratio of v$_{peak}$ to v$_{abs}$ approaching 0.6 around -10 days relative to \textit{B}-band maximum for the S II 5454 \AA{} line. Over time, the photosphere quickly recedes from the low-density material and the emitting region becomes more spherical, causing this ratio to approach 0 around \textit{B}-band maximum light. However, at early times the photosphere is very far out in the ejecta, so the emitting ejecta we observe must be at HVs, leading to a greater emission peak blueshift. Our \texttt{SYN++} spectrum supports this picture; the blueshift of the Si II 6355 \AA{} emission peak is proportional to the Si II ejecta velocity. Because SN\,2019ein has some of the highest absorption velocities measured at its earliest phases, the emission velocities of those lines are among the highest as well.

\subsection{Discussion}

We find that the high absorption and emission velocities at early times can be explained by an HV-only ejecta component, possibly due to mixing in an asymmetric explosion or optical depth effects in the outer layers of the ejecta. It is possible that both effects are at play; the models of \cite{Maeda2010} predict that the outer regions of the SN ejecta on the side opposite from an off-center ignition are less dense and produce HVG SNe. It could be that in this lower density environment, the ejecta is optically thinner, leading to a majority of the flux stemming from material moving along the observer's line of sight and producing blueshifted emission features. 

Another nearby SN Ia with a well-studied density structure is SN\,2012fr \citep{Childress2013,Maund2013,Contreras2018}.  \citet{Cain2018} found that SN\,2012fr showed signs of a shell-like density enhancement at low velocities, which could explain the unusual Si II velocity evolution as well as the presence of separate HV and PV features. However, SN\,2012fr and similar SNe Ia tend to be slow decliners ($\Delta m_{15} \lesssim 1$), HV yet LVG, and fall outside the BL region of the Branch diagram \citep{Contreras2018}. These classifications are at odds with those we present for SN\,2019ein. Therefore we suggest that SN\,2019ein most likely has a different density enhancement than the shell-like structure of SN\,2012fr.

One potential bias in our measurements is line blanketing. Line blanketing can warp the shape of the emission peaks, potentially biasing measurements of the peak wavelength. As noted by \citet{Branch2007}, synthetic spectra rarely if ever produce the significant emission peak Doppler shifts around maximum light observed by \citet{Blondin2006} and again here in SN\,2019ein. However, synthetic spectra also rarely produce only HV ejecta components, as we have done in our \texttt{SYN++} model. Furthermore, the observed emission shifts are not seen in just one line but globally, and seem to follow a similar evolution over time. This can be seen in the comparison between the B15 model spectra and the real data in the earliest epoch (Figure \ref{fig:synthspectra}). Therefore we conclude that line blanketing is unlikely able to reproduce the peculiar emission  blueshifts at all wavelengths and phases. 

Future observations will be necessary to provide more conclusive results on the geometry of the ejecta. For example, measuring a Doppler shift of the nebular-phase emission peaks could support the argument that SN\,2019ein has signatures of an aspherical explosion. In addition, early-time polarimetry data would provide an additional measurement of the asymmetries in both specific spectral features, such as the Si II absorption line, and globally via the continuum polarization. However, as discussed by \citet{Dessart2011} and \citet{Kasen2007}, low continuum polarization does not necessarily imply that the explosion was spherically symmetric; both authors find that even in models with significant asphericity, the line and continuum polarization could be low due to density and ionization effects. In the case of some geometries presented in \citet{Dessart2011} both an emission blueshift and a low polarization signal are produced, regardless of the underlying symmetry of the ejecta.

It is reasonable to question why SN\,2019ein is so unusual, even in a sample of other BL HV SNe. One possible explanation is early-time observations; it is possible that SN\,2019ein was first observed mere hours after explosion, allowing us to see the extremely high absorption and emission velocities at an earlier phase than other HV SNe. Another explanation is that SN\,2019ein was observed from a rare viewing angle, as would be the case if the explosion were strongly asymmetric. Either way, the analysis of early-time photometry and spectroscopy presented here demonstrates the importance of finding and observing SNe Ia quickly after explosion.
 
\section{Summary} \label{sec:conclusions}

We have presented photometric and spectroscopic observations of SN\,2019ein, a SN Ia with some of the highest early-time ejecta velocities ever measured. We observe a Si II 6355 \AA{} absorption velocity of 24,000 km s$^{-1}$ 14 days before \textit{B}-band maximum light. In addition, the early-time emission components of the P Cygni profiles appear blueshifted with respect to the host galaxy redshift, with emission peaks of Si II, Ca II, and S II moving at velocities up to or above 10,000 km s$^{-1}$. This emission blueshift is also among the highest ever measured, making SN\,2019ein an outlier even among other HV SNe. 

Radio observations taken as early as $<$4 days after explosion provide insight into the progenitor system of SN\,2019ein as well as the source of the HV ejecta. Our 3$\sigma$ VLA upper limits of 18, 25, and 23 $\mu$Jy at 3.87, 11.57, and 17.58 days after explosion are sensitive enough to rule out symbiotic progenitors for SN\,2019ein. We also rule out part of the parameter space of a single-degenerate model involving accretion from a main-sequence or slightly evolved companion at accretion rates $>3\times 10^{-7}$M$_{\odot}$ yr$^{-1}$, because the resulting fast optically thick winds would likely have created detectable circumstellar material. Such progenitor scenarios were also ruled out for the nearest and best-studied SNe Ia 2011fe and 2014J. Our upper limits cannot rule out models of a WD accreting at lower rates $\sim (1-3) \times 10^{-7}$ M$_{\odot}$ from a main-sequence or slightly evolved companion via winds that are sometimes interrupted by recurrent nova flashes. With our shell-interaction model \citep{Harris2016} we can rule out the presence of optically thin shells, which have been theoretically predicted to source HV ejecta, of masses $>10^{-6}$ M$_{\odot}$ at distances $<100$ AU from the progenitor. However, denser or more massive shells in the optically thick regime cannot be ruled out by the current model, and will be revisited in the future with a more sophisticated shell model that takes synchrotron self-absorption and radiative losses into account.

We find that SN\,2019ein is well fit by a delayed-detonation explosion model \citep{Blondin2015} except at early times, where our measured ejecta velocities are even higher than those predicted. By modeling the early spectra of SN\,2019ein, we find that both the high absorption and emission velocities may be due to a HV component of the ejecta that is detached from the photosphere. This detached component of the ejecta may be evidence of an aspherical distribution of intermediate-mass elements, perhaps due to mixing in an asymmetric explosion \citep{Seitenzahl}. Additionally, optical depth effects in the very outer layers of the ejecta may lead to an overall blueshift in the spectrum, as the majority of the flux observed comes from material moving along the observer's line of sight. These results highlight the need for more detailed modeling of SN ejecta, especially at early times. 

By studying a larger sample of HV SNe Ia, we can begin to probe the overlap between explosion models, asymmetries, and ejecta velocities. Results from such a sample would have implications on theories of Type Ia progenitor systems and explosion mechanisms. It is possible that a united picture will emerge, one in which the ejecta geometry and the viewing angle to a SN affect observables such as color, velocity, and light-curve width. A similar intrinsic difference has already been noted in the colors and host environments of HV and Normal SNe \citep{Wang2013,Zheng}, and may be used to reduce uncertainties in Type Ia distances, improving the precision of cosmological measurements.

\acknowledgments

We thank the anonymous referee for helpful feedback. C.P. thanks G. Mirek Brandt for helpful comments and discussions, Claudia Guti\'errez for generously sharing the velocity measurements and spectra used in Figures \ref{fig:gvels} and \ref{fig:gcomp}, Peter Il\'a\v{s} for his efforts in creating Figure \ref{fig:field}, and Chelsea Harris for guidance with her shell model and its application to the radio observations. C.P. is supported by NSF grant 1911225 and NASA Swift GI 1518168. S.K.S.\ and L.C.\ are supported by NSF grants AST-1412549, AST-1412980, and AST-1907790. D.J.S. is supported by NSF grants AST-1821967, 1821987, 1813708, 1813466, and 1908972. Research by S.V. is supported by NSF grant AST-1813176. M.S. is a Visiting Astronomer at the Infrared Telescope Facility, which is operated by the University of Hawaii under contract NNH14CK55B with the National Aeronautics and Space Administration.

The National Radio Astronomy Observatory is a facility of the National Science Foundation operated under cooperative agreement by Associated Universities, Inc. This research made use of observations from the LCO network, as well as the NASA/IPAC Extragalactic Database (NED) which is operated by the Jet Propulsion Laboratory, California Institute of Technology, under contract with NASA. 

\textit{\software}{lcogtsnpipe \citep{Valenti2016}, IRAF \citep{Tody1986,Tody1993}, SOUSA \citep{Brown2014}, CASA \citep[v5.4.1;][]{McMullin2007}, SYN++ \citep{Thomas}}

\appendix 

\section{Photometry Tables}
Here we present tables of the optical photometry (Table \ref{tab:optphot}), Swift UVOT photometry (Table \ref{tab:swiftphot}), and NIR photometry (Table \ref{tab:nirphot}) of SN\,2019ein. 

\begin{deluxetable}{lrrrrrr}[b]
\tablenum{4}
\tablecaption{Optical Photometry of SN\,2019ein\label{tab:optphot}}
\tablehead{
\colhead{MJD} & \colhead{U} & \colhead{B} & \colhead{V} & \colhead{g} & \colhead{r} & 
\colhead{i}}
\startdata
58607.9 & 15.6623$\pm$0.0178 & 16.2245$\pm$0.0068 & 16.0277$\pm$0.0094 & 16.0779$\pm$0.0069 & 16.0746$\pm$0.0083 & 16.3184$\pm$0.0123 \\
58607.9 & 15.6691$\pm$0.0182 & 16.1875$\pm$0.0077 & 16.0273$\pm$0.0095 & 16.0761$\pm$0.0067 & 16.0725$\pm$0.0083 & 16.3133$\pm$0.0133 \\
58609.9 & 14.9546$\pm$0.045 & 15.459$\pm$0.0246 & 15.3038$\pm$0.0066 & 15.2826$\pm$0.0049 & 15.2625$\pm$0.0055 & 15.4542$\pm$0.0074 \\
58609.9 & 15.2438$\pm$0.0428 & 15.3734$\pm$0.0089 & 15.3246$\pm$0.0058 & 15.2677$\pm$0.0048 & 15.25$\pm$0.0045 & 15.4743$\pm$0.0074 \\
58612.8 & 14.203$\pm$0.0131 & 14.6844$\pm$0.0071 & 14.6623$\pm$0.0059 & 14.5386$\pm$0.0163 & 14.5182$\pm$0.0064 & - \\
58612.8 & 14.1956$\pm$0.0125 & 14.69$\pm$0.0081 & 14.5789$\pm$0.0068 & - & 14.5123$\pm$0.0074 & - \\
\enddata
\tablecomments{This table is available in its entirety in machine-readable format.}
\end{deluxetable}

\begin{deluxetable}{lrrrrrr}[b!]
\tablenum{5}
\tablecaption{Swift UVOT Photometry of SN\,2019ein\label{tab:swiftphot}}
\tablehead{
\colhead{MJD} & \colhead{UVW2} & \colhead{UVM2} & \colhead{UVW1} & \colhead{U} & \colhead{B} & \colhead{V}}
\startdata
58606.5 & 19.924$\pm$0.342 & - & 18.398$\pm$0.173 & 16.675$\pm$0.09 & 16.735$\pm$0.074 & 16.37$\pm$0.096 \\
58607.8 & 18.801$\pm$0.16 & 20.359$\pm$0.35 & 17.559$\pm$0.11 & 15.727$\pm$0.068 & 16.103$\pm$0.063 & 15.698$\pm$0.073 \\
58608.8 & 18.3$\pm$0.122 & 20.517$\pm$0.36 & 16.903$\pm$0.083 & 15.143$\pm$0.057 & 15.728$\pm$0.059 & 15.474$\pm$0.067 \\
58609.8 & 18.069$\pm$0.111 & 20.292$\pm$0.302 & 16.512$\pm$0.074 & 14.816$\pm$0.051 & 15.242$\pm$0.049 & 15.159$\pm$0.063 \\
58612.6 & 17.119$\pm$0.09 & 18.836$\pm$0.208 & 15.603$\pm$0.06 & - & - & - \\
58616.1 & 16.71$\pm$0.076 & 18.099$\pm$0.118 & 15.321$\pm$0.06 & 13.855$\pm$0.043 & 14.189$\pm$0.042 & 14.037$\pm$0.049 \\
58617.4 & 16.667$\pm$0.069 & - & - & - & - & - \\
58617.4 & 16.642$\pm$0.078 & - & - & - & - & - \\
58619.2 & 16.701$\pm$0.091 & 17.665$\pm$0.146 & 15.329$\pm$0.061 & - & - & - \\
58620.2 & 16.681$\pm$0.074 & 17.795$\pm$0.093 & 15.423$\pm$0.06 & 14.096$\pm$0.044 & 14.152$\pm$0.042 & 13.843$\pm$0.047 \\
58629.4 & 17.814$\pm$0.07 & - & - & - & - & - \\
58629.4 & 17.568$\pm$0.105 & - & - & - & - & - \\ 
58630.3 & 17.796$\pm$0.097 & 18.718$\pm$0.149 & - & - & - & - \\
58702.2 & 20.162$\pm$0.347 & - & - & 18.175$\pm$0.173 & 17.536$\pm$0.094 & 16.956$\pm$0.119 \\
58712.1 & 20.262$\pm$0.361 & - & 19.144$\pm$0.246 & 18.219$\pm$0.166 & 17.86$\pm$0.104 & 17.12$\pm$0.122 
\enddata
\end{deluxetable}

\begin{deluxetable}{lrrr}[b!]
\tablenum{6}
\tablecaption{NIR Photometry of SN\,2019ein\label{tab:nirphot}}
\tablehead{
\colhead{MJD} & \colhead{J} & \colhead{H} & \colhead{K$_{s}$}}
\startdata
58606.8 & 15.97$\pm$0.20 & 15.40$\pm$0.31 & 15.31$\pm$0.29 \\
58606.9 & 15.91$\pm$0.22 & 15.56$\pm$0.28 & 15.32$\pm$0.32 \\
58612.9 & 14.50$\pm$0.17 & 14.61$\pm$0.23 & 14.41$\pm$0.27 \\
58617.8 & 14.29$\pm$0.24 & 14.50$\pm$0.24 & 14.24$\pm$0.30
\enddata
\end{deluxetable}


\begin{thebibliography}

\bibitem[Altavilla et al.(2007)]{Altavilla} Altavilla, G., Stehle, M., Ruiz-Lapuente, P., et al. \ 2007, \aap, 475, 585
\bibitem[Anderson et al.(2014)]{Anderson} Anderson, J.~P., Dessart, L., Gutierrez, C.~P., et al. \ 2014, \mnras, 441, 671
\bibitem[Arnett(1979)]{Arnett} Arnett, W.~D. \ 1979, \apjl, 230, L37
\bibitem[Bellm et al.(2019)]{Bellm} Bellm, E.~C., Kulkarni, S.~R., Graham, M.~J., et al.\ 2019, \pasp, 131, 018002
\bibitem[Benetti et al.(2004)]{Benetti2004} Benetti, S., Meikle, P., Stehle, M., et al. \ 2004, \mnras, 348, 261
\bibitem[Benetti et al.(2005)]{Benetti2005} Benetti, S., Cappellaro, E., Mazzali, P.~A., et al. \ 2005, \apj, 623, 1011
\bibitem[Betoule et al.(2014)]{Betoule} Betoule, M., Kessler, R., Guy, J., et al. \ 2014, \aap, 568, A22
\bibitem[Blondin et al.(2006)]{Blondin2006} Blondin, S., Dessart, L., Leibundgut, B., et al. \ 2006, \aj, 131, 1648
\bibitem[Blondin et al.(2012)]{Blondin2012} Blondin, S., Matheson, T., Kirshner, R.~P., et al. \ 2012, \aj, 143, 126
\bibitem[Blondin, Dessart, \& Hillier(2015)]{Blondin2015} Blondin, S., Dessart, L., \& Hillier, D.~J. \ 2015, \mnras, 448, 2766
\bibitem[Bloom et al.(2012)]{Bloom} Bloom, J.~S., Kasen, D., Shen, K.~J., et al.\ 2012, \apjl, 744, L17
\bibitem[Branch et al.(2006)]{Branch2006} Branch, D., Dang, L.~C., Hall, N., et al. \ 2006, PASP, 118, 560
\bibitem[Branch et al.(2007)]{Branch2007} Branch, D., Troxel, M.~A., Jeffery, D.~J., et al. \ 2007, \pasp, 119, 709
\bibitem[Breeveld et al.(2011)]{Breeveld} Breeveld, A.~A., Landsman, W., Holland, S.~T., et al. \ 2011, in AIP Conf. Ser. 1358, AIP Conf. Ser., ed. J. E. McEnery, J. L. Racusin, \& N. Gehrels (Melville, NY: AIP), 373
\bibitem[Brown et al.(2013)]{Brown} Brown, T.~M., Baliber, N., Bianco, F.~B., et al. \ 2013, \pasp, 125, 1031
\bibitem[Brown et al.(2014)]{Brown2014} Brown, P.~J., Breeveld, A.~A., Holland, S., et al. \ 2014, \apss, 354, 89
\bibitem[Brown et al.(2018)]{Brown2018} Brown, P.~J., Perry, J.~M., Beeny, B.~A., et al.\ 2018, \apj, 867, 56
\bibitem[Bulla et al.(2016)]{Bulla2016} Bulla, M., Sim, S.~A., Kromer, M., et al. \ 2016, \mnras, 462, 1039
\bibitem[Burke et al.(2019)]{Burke} Burke, J., Arcavi, I., Howell, D.~A., et al. \ 2019, ATel, 12719, 1
\bibitem[Cain et al.(2018)]{Cain2018} Cain, C., Baron, E., Phillips, M.~M., et al.\ 2018, \apj, 869, 162
\bibitem[Cendes et al.(2020)]{Cendes2020} Cendes, Y., Drout, M.~R., Chomiuk, L., et al.\ 2020, \apj, 894, 39
\bibitem[Chen et al.(2011)]{Chen2011} Chen, X., Han, Z., \& Tout, C.~A.\ 2011, \apjl, 735, L31
\bibitem[Chevalier(1982)]{Chevalier1982} Chevalier, R.~A.\ 1982, \apj, 259, 302
\bibitem[Chevalier(1984)]{Chevalier1984} Chevalier, R.~A.\ 1984, \apjl, 285, L63
\bibitem[Chevalier(1998)]{Chevalier1998} Chevalier, R.~A.\ 1998, \apj, 499, 810
\bibitem[Chevalier \& Fransson(2006)]{Chevalier2006} Chevalier, R.~A., \& Fransson, C.\ 2006, \apj, 651, 381
\bibitem[Childress et al.(2013)]{Childress2013} Childress, M.~J., Scalzo, R.~A., Sim, S.~A., et al.\ 2013, \apj, 770, 29
\bibitem[Childress et al.(2014)]{Childress} Childress, M.~J., Filippenko, A.~V., Ganeshalingam, M., et al. \ 2014, \mnras, 437, 338
\bibitem[Chomiuk et al.(2012)]{Chomiuk2012} Chomiuk, L., Soderberg, A.~M., Moe, M., et al.\ 2012, \apj, 750, 164
\bibitem[Chomiuk et al.(2016)]{Chomiuk2016} Chomiuk, L., Soderberg, A.~M., Chevalier, R.~A., et al.\ 2016, \apj, 821, 119 
\bibitem[Cikota et al.(2019)]{Cikota} Cikota, A., Patat, F., Wang, L., et al.\ 2019, \mnras, 490, 578
\bibitem[Contreras et al.(2018)]{Contreras2018} Contreras, C., Phillips, M.~M., Burns, C.~R., et al.\ 2018, \apj, 859, 24
\bibitem[Cornwell et al.(2008)]{Cornwell2008} Cornwell, T.~J., Golap, K., \& Bhatnagar, S.\ 2008, IEEE Journal of Selected Topics in Signal Processing, 2, 647
\bibitem[Dessart \& Hillier(2005)]{Dessart2005} Dessart, L. \& Hillier, D.~J. \ 2005, \aap, 439, 671
\bibitem[Dessart \& Hillier(2011)]{Dessart2011} Dessart, L. \& Hillier, D.~J. \ 2011, \mnras, 415, 3497
\bibitem[Dessart et al.(2014)]{Dessart2014} Dessart, L., Blondin, S., Hillier, D.~J., et al. \ 2013, \mnras, 441, 532
\bibitem[Filippenko et al.(1992b)]{Filippenko1992} Filippenko, A.~V., Richmond, M.~W., Branch, D., et al. \ 1992, \aj, 104, 1543
\bibitem[Filippenko et al.(1992a)]{Filippenko1992a} Filippenko, A.~V., Richmond, M.~W., Matheson, T., et al.\ 1992, \apjl, 384, L15
\bibitem[Fink et al.(2010)]{Fink2010} Fink, M., R{\"o}pke, F.~K., Hillebrandt, W., et al.\ 2010, \aap, 514, A53
\bibitem[Folatelli et al.(2012)]{Folatelli2012} Folatelli, G., Phillips, M.~M., Morrell, N., et al.\ 2012, \apj, 745, 74 
\bibitem[Foley \& Kasen(2011)]{Foley} Foley, R.~J. \& Kasen, D. \ 2011, \apj, 729, 55F
\bibitem[Ganeshalingam et al.(2011)]{Ganeshalingam} Ganeshalingam, M., Li, W., \& Filippenko, A.~V.\ 2011, \mnras, 416, 2607
\bibitem[Garavini et al.(2007)]{Garavini} Garavini, G., Folatelli, G., Nobili, S., et al. \ 2007, \aap, 470, 411
\bibitem[Gehrels et al.(2004)]{Gehrels} Gehrels, N., Chincarini, G., Giommi, P., et al.\ 2004, \apj, 611, 1005
\bibitem[Gerardy et al.(2004)]{Gerardy} Gerardy, C.~L., H{\"o}flich, P., Fesen, R.~A., et al. \ 2004, \apj, 607, 391
\bibitem[Gutierrez et al.(2016)]{Gutierrez} Guti{\'e}rrez, C.~P., Gonz{\'a}lez-Gait{\'a}n, S., Folatelli, G., et al. \ 2016, \aap, 590A, 5
\bibitem[Guy et al.(2007)]{Guy} Guy, J., Astier, P., Baumont, S., et al. \ 2007, \aap, 466, 11
\bibitem[Guy et al.(2010)]{Guy2010} Guy, J., Sullivan, M., Conley, A., et al. \ 2010, \aap, 523A, 7
\bibitem[Hachisu et al.(1999)]{Hachisu1999} Hachisu, I., Kato, M., \& Nomoto, K.\ 1999, \apj, 522, 487 
\bibitem[Hamuy et al.(1996)]{Hamuy} Hamuy, M., Phillips, M.~M., Suntzeff, N.~B., et al.\ 1996, \aj, 112, 2391
\bibitem[Harris et al.(2016)]{Harris2016} Harris, C.~E., Nugent, P.~E., \& Kasen, D.~N.\ 2016, \apj, 823, 100
\bibitem[Hillier \& Miller(1998)]{Hillier} Hillier, D.~J. \& Miller, D.~L. \ 1998, \apj, 496, 407
\bibitem[Horesh et al.(2012)]{Horesh2012} Horesh, A., Kulkarni, S.~R., Fox, D.~B., et al.\ 2012, \apj, 746, 21
\bibitem[Howell et al.(2009)]{Howell2009} Howell, D.~A., Sullivan, M., Brown, E.~F., et al. \ 2009, \apj, 691, 661
\bibitem[Hsiao et al.(2013)]{Hsiao2013} Hsiao, E.~Y., Marion, G.~H., Phillips, M.~M., et al. \ 2013, \apj, 766, 72
\bibitem[Hsiao et al.(2015)]{Hsiao2015} Hsiao, E.~Y., Burns, C.~R., Contreras, C., et al. \ 2015, \aap, 578, A9
\bibitem[Hsiao et al.(2019)]{Hsiao2019} Hsiao, E.~Y., Phillips, M.~M., Marion, G.~H., et al.\ 2019, \pasp, 131, 014002
\bibitem[Iben \& Tutukov(1984)]{Iben1984} Iben, Jr., I. \& Tutukov, A.~V. \ 1984, \apjs, 54, 335
\bibitem[Iwamoto et al.(1999)]{Iwamoto} Iwamoto, K., Brachwitz, F., \& Nomoto, K. \ 1999, \apjs, 125, 439
\bibitem[Kasen \& Plewa(2007)]{Kasen2007} Kasen, D. \& Plewa, T. \ 2007, \apj, 662, 459
\bibitem[Kawabata et al.(2020)]{Kawabata} Kawabata, M., Maeda, K., Yamanaka, M., et al. \ 2020, \apj, 893, 143
\bibitem[Khokhlov(1991)]{Khokhlov} Khokhlov, A.~M. \ 1991, \aap, 245, 114
\bibitem[K{\"o}nyves-T{\'o}th et al.(2020)]{Konyves-Toth} K{\"o}nyves-T{\'o}th, R., Vink{\'o}, J., Ordasi, A., et al.\ 2020, \apj, 892, 121
\bibitem[Kromer et al.(2010)]{Kromer2010} Kromer, M., Sim, S.~A., Fink, M., et al.\ 2010, \apj, 719, 1067
\bibitem[Landolt(1992)]{Landolt} Landolt, A.~U.\ 1992, \aj, 104, 340
\bibitem[Macaulay et al.(2019)]{Macaulay} Macaulay, E., Nichol, R.~C., Bacon, D., et al. \ 2019, \mnras, 486, 2184
\bibitem[Maeda et al.(2010)]{Maeda2010} Maeda, K., Benetti, S., Stritzinger, M., et al. \ 2010, \nat, 466, 82
\bibitem[Maguire et al.(2014)]{Maguire} Maguire, K., Sullivan, M., Pan, Y.~C., et al. \ 2014, \mnras, 444, 3258
\bibitem[Maund et al.(2010)]{Maund} Maund, J.~R., H{\"o}flich, P., Patat, F., et al. \ 2010, \apjl, 725, L167
\bibitem[Maund et al.(2013)]{Maund2013} Maund, J.~R., Spyromilio, J., H{\"o}flich, P.~A., et al. \ 2013, \mnras, 433, L20
\bibitem[Mazzali et al.(2005)]{Mazzali} Mazzali, P.~A., Benetti, S., Stehle, M., et al. \ 2005, \mnras, 357, 200
\bibitem[McMullin et al.(2007)]{McMullin2007} McMullin, J.~P., Waters, B., Schiebel, D., et al.\ 2007, adass XVI, ASP Conf. Ser. 376, ed. R. A. Shaw, F. Hill, \& D. J. Bell (San Francisco, CA: ASP), 127
\bibitem[Milne et al.(2013)]{Milne2013} Milne, P.~A., Brown, P.~J., Roming, P.~W.~A., et al.\ 2013, \apj, 779, 23
\bibitem[Mulligan \& Wheeler(2017)]{Mulligan2017} Mulligan, B.~W. \& Wheeler, J.~C. \ 2017, \mnras, 467, 778
\bibitem[Mulligan, \& Wheeler(2018)]{Mulligan2018} Mulligan, B.~W., \& Wheeler, J.~C.\ 2018, \mnras, 476, 1299
\bibitem[Nagao et al.(2019)]{Nagao} Nagao, T., Cikota, A., Patat, F., et al.\ 2019, \mnras, 489, L69
\bibitem[Nomoto et al.(2013)]{Nomoto2013} Nomoto, K., Kamiya, Y., \& Nakasato, N. \ 2013, IAU Symposium, 281, ``Binary Paths to Type Ia Supernova Explosions," ed. R. Di Stefano, M. Orio, \& M. Moe (Cambridge: Cambridge University Press), 253
\bibitem[Panagia et al.(2006)]{Panagia2006} Panagia, N., Van Dyk, S.~D., Weiler, K.~W., et al.\ 2006, \apj, 646, 369
\bibitem[Parrent, Friesen, \& Parthasarathy(2014)]{Parrent} Parrent, J., Friesen, B., \& Parthasarathy, M. \ 2014, \apss, 351, 1P
\bibitem[Parrent et al.(2012)]{Parrent2012} Parrent, J.~T., Howell, D.~A., Friesen, B., et al.\ 2012, \apjl, 752, L26
\bibitem[Parrent et al.(2011)]{Parrent2011} Parrent, J.~T., Thomas, R.~C., Fesen, R.~A., et al. \ 2011, \apj, 732, 30
\bibitem[Patat et al.(2011)]{Patal2011} Patat, F., Chugai, N.~N., Podsiadlowski, P., et al.\ 2011, \aap, 530, A63
\bibitem[Pereira et al.(2013)]{Pereira} Pereira, R., Thomas, R.~C., Aldering, G., et al. \ 2013, \aap, 554, A27
\bibitem[Perlmutter et al.(1999)]{Perlmutter1999} Perlmutter, S., Aldering, G., Goldhaber, G., et al. \ 1999, \apj, 517, 565
\bibitem[Phillips et al.(1992)]{Phillips1992} Phillips, M.~M., Wells, L.~A., Suntzeff, N.~B., et al. \ 1992, \aj, 103, 1632
\bibitem[Phillips(1993)]{Phillips} Phillips, M.~M. \ 1993, \apj, 413L, 105P 
\bibitem[Rau \& Cornwell(2011)]{Rau2011} Rau, U., \& Cornwell, T.~J.\ 2011, \aap, 532, A71 
\bibitem[Rayner et al.(2003)]{Rayner} Rayner, J.~T., Toomey, D.~W., Onaka, P.~M., et al. \ 2003, \pasp, 115, 362
\bibitem[Riess et al.(1998)]{Riess1998} Riess, A.~G., Filippenko, A.~V., Challis, P., et al. \ 1998, \aj, 116, 1009
\bibitem[Riess et al.(2019)]{Riess2019} Riess, A.~G., Casertano, S., Yuan, W., et al. \ 2019, \apj, 876, 85
\bibitem[Ryder et al.(2004)]{Ryder2004} Ryder, S.~D., Sadler, E.~M., Subrahmanyan, R., et al.\ 2004, \mnras, 349, 1093
\bibitem[Salas et al.(2013)]{Salas2013} Salas, P., Bauer, F.~E., Stockdale, C., et al.\ 2013, \mnras, 428, 1207
\bibitem[Schlafly \& Finkbeiner(2011)]{Schlafly} Schlafly, E.~F. \& Finkbeiner, D.~P. \ 2011, \apj, 737, 103
\bibitem[Schmidt et al.(1998)]{Schmidt1998} Schmidt, B.~P., Suntzeff, N.~B., Phillips, M.~M., et l. \ 1998, \apj, 507, 46
\bibitem[SDSS Collaboration et al.(2017)]{SDSS} SDSS Collaboration, Albareti, F.~D., Allende Prieto, C., et al. \ 2017, \apjs, 233, 25
\bibitem[Seaquist \& Taylor(1990)]{Seaquist1990} Seaquist, E.~R., \& Taylor, A.~R.\ 1990, \apj, 349, 313
\bibitem[Seaquist et al.(1993)]{Seaquist1993} Seaquist, E.~R., Krogulec, M., \& Taylor, A.~R.\ 1993, \apj, 410, 260
\bibitem[Seitenzahl et al.(2013)]{Seitenzahl} Seitenzahl, I.~R., Ciaraldi-Schoolmann, F., R{\"o}pke, F.~K., et al. \ 2013, \mnras, 429, 1156
\bibitem[Shen \& Bildsten(2007)]{Shen2007} Shen, K.~J., \& Bildsten, L.\ 2007, \apj, 660, 1444 
\bibitem[Silverman et al.(2012)]{Silverman} Silverman, J.~M., Foley, R.~J., Filippenko, A.~V., et al.\ 2012, \mnras, 425, 1789
\bibitem[Silverman \& Filippenko(2012)]{Silverman2012} Silverman, J.~M., \& Filippenko, A.~V.\ 2012, \mnras, 425, 1917 
\bibitem[Sironi \& Spitkovsky(2011)]{Sironi2011} Sironi, L., \& Spitkovsky, A.\ 2011, \apj, 726, 75
\bibitem[Skrutskie et al.(2006)]{Skrutskie} Skrutskie, M.~F., Cutri, R.~M., Stiening, R., et al.\ 2006, \aj, 131, 1163
\bibitem[Soderberg et al.(2006)]{Soderberg2006} Soderberg, A.~M., Chevalier, R.~A., Kulkarni, S.~R., et al.\ 2006, \apj, 651, 1005
\bibitem[Soderberg et al.(2012)]{Soderberg2012} Soderberg, A.~M., Margutti, R., Zauderer, B.~A., et al.\ 2012, \apj, 752, 78
\bibitem[Stetson(1987)]{Stetson} Stetson, P.~B.\ 1987, \pasp, 99, 191
\bibitem[Stritzinger et al.(2006)]{Stritzinger2006} Stritzinger, M., Mazzali, P.~A., Sollerman, J., et al. \ 2006, \aap, 460, 793
\bibitem[Stritzinger et al.(2018)]{Stritzinger2018} Stritzinger, M.~D., Shappee, B.~J., Piro, A.~L., et al.\ 2018, \apjl, 864, L35
\bibitem[Thomas, Nugent, \& Meza(2011)]{Thomas} Thomas, R.~C., Nugent, P.~E., \& Meza, J.~C. \ 2011, \pasp, 123, 237
\bibitem[Tody(1986)]{Tody1986} Tody, D.\ 1986, \procspie, 627, ``The IRAF Data Reduction and Analysis System," ed. D. L. Crawford, 733
\bibitem[Tody(1993)]{Tody1993} Tody, D.\ 1993, adass, 52, ``IRAF in the Nineties," ed. R. J. Hanisch, R. J. V. Brissenden, \& J. Barnes, 173
\bibitem[Tonry et al.(2019)]{Tonry} Tonry, J., Denneau, L., Heinze, A., et al. \ 2019, TNSTR, 2019-678, 1
\bibitem[P{\'e}rez-Torres et al.(2014)]{Torres2014} P{\'e}rez-Torres, M.~A., Lundqvist, P., Beswick, R.~J., et al.\ 2014, \apj, 792, 38
\bibitem[Valenti et al.(2016)]{Valenti2016} Valenti, S., Howell, D.~A., Stritzinger, M.~D., et al. \ 2016, \mnras, 459, 3939
\bibitem[Wang \& Wheeler(2008)]{WangWheeler} Wang, L. \& Wheeler, J.~C. \ 2008, \araa, 46, 433
\bibitem[Wang et al.(2009)]{Wang2009} Wang, X., Filippenko, A.~V., Ganeshalingam, M., et al. \ 2009, \apjl, 699, L139
\bibitem[Wang et al.(2013)]{Wang2013} Wang, X., Wang, L., Filippenko, A.~V., et al. \ 2013, Science, 340, 170
\bibitem[Weiler et al.(2007)]{Weiler2007} Weiler, K.~W., Williams, C.~L., Panagia, N., et al.\ 2007, \apj, 671, 1959
\bibitem[Whelan \& Iben(1973)]{Whelan1973} Whelan, J., \& Iben, Jr., I. \ 1973, \apj, 186, 1007
\bibitem[Wood-Vasey, \& Sokoloski(2006)]{WoodVasey2006} Wood-Vasey, W.~M., \& Sokoloski, J.~L.\ 2006, \apjl, 645, L53
\bibitem[Woosley \& Kasen(2011)]{Woosley2011} Woosley, S.~E., \& Kasen, D.\ 2011, \apj, 734, 38
\bibitem[Zheng, Kelly, \& Filippenko(2018)]{Zheng} Zheng, W., Kelly, P.~L., \& Filippenko, A.~V. \ 2018, \apj, 858, 104

\end{thebibliography}
\end{document}